\documentstyle[11pt,epsf]{article}
 1
 1
 1

\def\lsim{\mathrel{\raise.3ex\hbox{$<$\kern-.75em\lower1ex\hbox{$\sim$}}}}
\def\gsim{\mathrel{\raise.3ex\hbox{$>$\kern-.75em\lower1ex\hbox{$\sim$}}}}
\begin{document}
\noindent
\thispagestyle{empty}
\renewcommand{\thefootnote}{\fnsymbol{footnote}}
\begin{flushright}
{\bf UCSD/PTH 97-04}\\
{\bf hep-ph/9702331}\\
{\bf February 1997}\\
\end{flushright}
\vspace{.5cm}
\begin{center}
  \begin{Large}\bf
The Vacuum Polarization Function to ${\cal{O}}(\alpha^2)$ Accuracy
\\[1mm]
Near Threshold and Darwin Corrections
  \end{Large}
  \vspace{1.5cm}

\begin{large}
 A.H. Hoang
\end{large}
\begin{center}
\begin{it}
   Department of Physics,
   University of California, San Diego,\\
   La Jolla, CA 92093--0319, USA\\ 
\end{it} 
\end{center}

  \vspace{4cm}
  {\bf Abstract}\\
\vspace{0.3cm}
%
\noindent
\begin{minipage}{15.0cm}
\begin{small}
The QED vacuum polarization function is calculated to ${\cal{O}}(\alpha^2)$
(next-to-leading order)
accuracy in the threshold regime by using the concept of effective
field theories to resum
diagrams with the instantaneous Coulomb exchange of longitudinally
polarized photons.
It is shown that the ${\cal{O}}(\alpha^2)$ contributions
are of order $\alpha^2$ in size rather than $\alpha^2/\pi^2$.
The vacuum polarization contributions to the ${\cal{O}}(\alpha^6)$
hyperfine splitting
of the positronium ground state are recalculated and differences with 
an older calculation are pointed out. 
The results are used to determine ${\cal{O}}(C_F^2\alpha_s^2)$ 
(next-to-next-to-leading order) Darwin corrections to heavy quark-antiquark 
bound state $l=0$ wave functions at the
origin and to the heavy quark-antiquark production cross section in
$e^+e^-$ collisions in the threshold region. The absolute value 
of the corrections amounts to $10\%-20\%$ and $17\%-34\%$
in the modulus squared of the ground state wave functions at the origin for 
the $b\bar b$ and $c\bar c$ systems, respectively. In the case of 
the $t\bar t$ production cross section in the threshold region the
absolute value of the corrections is between $2\%$ and $6\%$ around
the $1S$ peak and between $1\%$ and $2\%$  for higher energies.
A critical comment on recent QCD sum rule calculations for the $\Upsilon$ 
system is made.
\end{small}
\end{minipage}
\end{center}
\setcounter{footnote}{0}
\renewcommand{\thefootnote}{\arabic{footnote}}
\vspace{1.2cm}
%
%
%
\newpage
\noindent
\section{Introduction}
\label{sectionintroduction}
In recent years many sophisticated methods have been developed to calculate
higher order (``multi-loop'') QCD radiative corrections for high
energy quantities for which it is believed that an expansion in terms
of Feynman diagrams with a certain number of loops represents an
excellent approximation to the predictions of quantum
chromodynamics. Notable examples are the hadronic cross section
in $e^+e^-$~collisions at LEP energies or the (photonic) vacuum
polarization function. In the high energy limit, where the 
quarks can be treated as massless, these quantities have been
calculated up to three loops~\cite{Chet1,Gorishny1,Surguladze1,Chet2}. 
However, future experiments (NLC,
B-factory and $\tau$-charm factory) will test the vacuum polarization
function and the hadronic cross section also in the kinematic regime
close to heavy quark-antiquark thresholds, where bound state effects
become important. The threshold regime is characterized by the
relation 
\begin{equation}
|\beta|\, \lsim\, \alpha_s \,, \quad 
\beta \, \equiv \, \sqrt{1-4\,\frac{M_Q^2}{q^2+i\epsilon}} \,,
\end{equation}
where $M_Q$ is the heavy quark mass and $\sqrt{q^2}$ denotes the
c.m. energy. In the process of heavy quark-antiquark production above
threshold, $q^2 > 4 M_Q^2$, $\beta$ is equal to the velocity of the
quarks in the c.m. frame. We therefore call $\beta$ ``velocity'' in
the remainder of this work, even if $q^2 < 4 M_Q^2$. In the
threshold regime the accuracy of theoretical predictions to the
hadronic cross section and to the vacuum polarization function is much
poorer than for high energies. Aside from definitely non-perturbative
effects (in the sense of ``not calculable from first principles in
QCD''), the breakdown of the perturbative expansion in the number of
loops makes any theoretical description in the threshold region
difficult. This breakdown of the perturbation series is indicated by
power ($1/\beta$) or logarithmic ($\ln\beta$) divergences in the
velocity which blow up if evaluated very close to the threshold
point. Some of these divergences ({\it e.g.} the $\alpha^n/\beta^n$, $n>1$, 
Coulomb singularities
in the Dirac form factor $F_1$ describing the electromagnetic vertex)
can be treated by using well-known results from non-relativistic
quantum mechanics, but a systematic way to calculate higher-order
corrections in the threshold regime seems to be far from obvious, at
least from the point of view of covariant perturbation theory in 
the number of loops. This type of perturbation theory will 
be referred to as ``conventional perturbation theory'' from now on
in this work. 
\par
On the other hand, there are many examples of heavy 
quark-antiquark  bound state properties where the complete knowledge
of higher-order corrections would be extremely valuable. Most of the
present analyses (see {\it e.g.}~\cite{Bodwin1}) are based on leading and
next-to-leading order calculations. Here, higher-order corrections
could significantly increase the precision of present theoretical
calculations, but could also serve as an instrument to test how
trustworthy certain theoretical predictions are and to estimate the
size of theoretical uncertainties. Further, they could contribute
toward a better understanding of the role of non-perturbative effects
(in the sense mentioned above) in apparent discrepancies between the 
determination of the size of the strong coupling from the $\Upsilon(1S)$ 
decay rates~\cite{Hinchliffe1} and QCD sum rule calculations for the
$\Upsilon$ system~\cite{Voloshin1}\footnote{
During completion of this paper we became aware of a new publication, 
where QCD sum rules for the $\Upsilon$ system are used to determine the
strong coupling and the bottom quark mass~\cite{Pich1}. 
We will give a brief comment on this publication and 
on~\cite{Voloshin1} at the end of Section~\ref{SectionThreshold}.
} 
on the one hand, 
and from the LEP experiments on the other.
\par
The framework in which bound state properties and also dynamical
quantities in the threshold regime can be calculated in a systematic
way to arbitrary precision is non-relativistic quantum chromodynamics
(NRQCD)~\cite{Bodwin1} which is based on the concept of effective field
theories. In the kinematic regime where bound states occur
and slightly above the threshold, NRQCD is superior to conventional
perturbation theory in QCD and (at least from the practical point of view)
also to the Bethe-Salpeter approach, because it allows for an easy
and transparent separation of long- and short-distance physics
contributions. This is much more difficult and cumbersome
with the former two methods. However, we would like to emphasize that
all methods lead to the same results. As an effective field theory,
NRQCD needs the input from short-distance QCD in order to produce
viable predictions in accordance with quantum chromodynamics. This
adjustment of NRQCD to QCD is called the {\it matching procedure} and
generally requires multi-loop calculations in the framework of
conventional perturbation theory at the level of the intended
accuracy.
\par
In this work we demonstrate the efficient use of the concept of
effective field theories to calculate the QED vacuum
polarization function 
in the threshold region to ${\cal{O}}(\alpha^2)$ accuracy. In order to
convince the reader of the simplicity of the approach we use our
result to recalculate the vacuum polarization contributions to the
${\cal{O}}(\alpha^6)$ hyperfine splitting of the positronium ground
state energy level without referring back to the Bethe-Salpeter
equation. Differences between our result and an older 
calculation~\cite{Barbieri1,Barbieri2}, are pointed out. We analyse the vacuum
polarization function at the bound state energies and above threshold
and, in particular, concentrate on the size of the
${\cal{O}}(\alpha^2)$ corrections. It is shown that the size of the
${\cal{O}}(\alpha^2)$ corrections in the threshold regime is of order
$\alpha^2$ rather than $\alpha^2/\pi^2$ which is  a consequence of their
long-distance origin. 
In a second step our results for the QED vacuum polarization function
are applied to calculate
${\cal{O}}(C_F^2\alpha_s^2)$ (next-to-next-to-leading order) Darwin
corrections to the heavy quark-antiquark $l=0$ bound state
wave functions at the origin and to 
the cross section of heavy quark-antiquark production
in $e^+e^-$ annihilation (via a virtual photon) in the threshold
region. The corresponding unperturbed quantities are the
solutions of the Schr\"odinger equation for a stable quark-antiquark
pair with a Coulomb-like QCD potential, 
$V_{\mbox{\tiny QCD}}(r)=-C_F\alpha_s/r$. It is demonstrated that the
size of the ${\cal{O}}(C_F^2\alpha_s^2)$ Darwin corrections is also of order 
$\alpha_s^2$ rather than $\alpha_s^2/\pi^2$. We present
simple physical arguments that the scale of the strong coupling
governing the ${\cal{O}}(C_F^2\alpha_s^2)$ Darwin corrections is of order
$C_F M_Q \alpha_s$ and we analyze the size of the corrections for
the $t\bar t$, $b\bar b$ and $c\bar c$ systems assuming that
the size of the Darwin corrections can be taken
as an order of magnitude estimate for all (yet unknown) 
${\cal{O}}(\alpha_s^2)$ corrections. The sign
of the latter  corrections and their actual numerical values can, of course,
only be determined by an explicit calculation of all
${\cal{O}}(\alpha_s^2)$ contributions.
\par 
At this point we want to emphasize that our approach does not depend
on any model-like assumptions, but represents a first principles QCD
calculation. The only assumptions (for heavy quarks)
are that (i) the instantaneous ({\it i.e.} uncrossed)
Coulomb-like exchange of longitudinal gluons (in Coulomb gauge)
between the heavy quarks leads to the dominant contributions in the
threshold regime and is the main reason for heavy quark-antiquark
bound state formation and that (ii) all further interactions can be
treated as a perturbation. We believe that the actual size of the
${\cal{O}}(\alpha_s^2)$ corrections can then serve as an important
{\it a posteriori} justification or falsification of these assumptions
for the different heavy quark-antiquark systems. Finally, we
address the question whether bound state effects can lead to large
corrections to the vacuum polarization function in kinematical regions
far from the actual threshold
regime. We come to the conclusion that such corrections
do not exist.
\par
The program for this work is organized as follows: In
Section~\ref{SectionCalculation} the calculation of the QED vacuum
polarization function to ${\cal{O}}(\alpha^2)$ accuracy in the
threshold region is presented. We define a
renormalized version of the Coulomb Green function for zero distances,
which allows for application of (textbook quantum mechanics) 
time-independent perturbation theory to determine higher-order 
corrections to 
wave functions and energy levels. For completeness we also give an
expression for the QED vacuum polarization function valid for all
energies with ${\cal{O}}(\alpha^2)$ accuracy. In
Section~\ref{SectionAnalysis} the QED vacuum polarization function in
the threshold region is analysed with special emphasis on the size of
the ${\cal{O}}(\alpha^2)$ corrections, and the 
${\cal{O}}(\alpha^6)$ vacuum polarization contributions to the
positronium ground state hyperfine splitting are calculated. 
Section~\ref{SectionQCD} is devoted to the determination and
analysis of the ${\cal{O}}(C_F^2\alpha_s^2)$ Darwin corrections to the
bound state wave functions at the origin and the production
cross section in the threshold regime for the different heavy
quark-antiquark systems. In Section~\ref{SectionThreshold} we comment
on the existence of threshold effects far from threshold and
Section~\ref{SectionSummary} contains a summary. 
\par
\vspace{0.5cm}
\section{Determination of the QED Vacuum Polarization Function \\
in the Threshold Region}
\label{SectionCalculation}
We consider the QED vacuum polarization function $\Pi$ defined through
the one-particle-irreducible current-current correlator
\begin{equation}
\big(\,q^2\,g^{\mu\nu}\,-\,q^\mu q^\nu\,\big)\,\Pi(q^2) \, \equiv \, 
-i \int \!{\rm d}^4x\,e^{i\,qx}\,
   \langle 0|\,T\,j^\mu(x)\,j^\nu(0)\,|0 \rangle
\,,
\end{equation}
where $j^\mu(x)=i e \bar{\Psi}(x)\gamma^\mu\Psi(x)$ denotes the 
electromagnetic current. $\Psi$ represents
the Dirac field of the electron with charge $e$. 
According to the standard subtraction procedure, 
$\Pi$ vanishes for $q^2=0$. It has been shown
in~\cite{Braun1} that in the kinematical region close to the $e^+e^-$
threshold point $q^2=4\,M^2$, where $M$ denotes the electron mass, the
current-current correlator $\Pi$ is directly related to the
Green function
$G^0_E(\vec x,\vec{x}^\prime)$ of the positronium Schr\"odinger
equation
\begin{equation}
\bigg[\,
-\frac{1}{M}\vec{\nabla}^2_{\vec x} - \frac{\alpha}{|\vec x|} - E
\,\bigg]\,G^0_E(\vec{x},\vec{x}^\prime) \, = \,
\delta^{(3)}(\vec{x}-\vec{y})
\,,
\end{equation}
where $E$ denotes the energy relative to the threshold point,
$E\equiv\sqrt{q^2}-2M$, and $\alpha$ is the fine structure constant. 
Explicit analytic expressions for the Green function have been
calculated in a number of classical papers~\cite{Wichmann1}.
The proper relation between the vacuum
polarization function and the Green function in the threshold region
reads~\cite{Braun1} 
\begin{equation}
\Pi_{Thr}^{0,{\cal{O}}(\alpha^2)}(q^2) \, = \, 
\frac{8\,\alpha\,\pi}{q^2}\,G^0_E(0,0).
\label{PGunrenormalized}
\end{equation}
Because we are only interested in ${\cal{O}}(\alpha^2)$ accuracy in
the threshold region we effectively can replace the factor $1/q^2$ in
eq.~(\ref{PGunrenormalized}) by $1/4M^2$. 
\par
For illustration, let us now examine the one- and two-loop
contributions to the vacuum polarization function and the expression
for the vacuum polarization function from non-relativistic quantum
mechanics according to eq.(\ref{PGunrenormalized}). 
The one- and two-loop contributions to $\Pi$ have been known for
quite a long time for all energy and mass 
assignments~\cite{Kallensabry1,Schwinger1, Barbieri3}. Far 
from the threshold point those loop results provide an excellent
approximation to the QED vacuum polarization function at the
${\cal{O}}(\alpha^2)$ accuracy level,
\begin{equation}
\Pi^{\mbox{\tiny 2 loop}}_{\mbox{\tiny QED}}(q^2) \, = \, 
\Big(\frac{\alpha}{\pi}\Big)\,\Pi^{(1)}(q^2) 
\, + \,
\Big(\frac{\alpha}{\pi}\Big)^2\,\Pi^{(2)}(q^2) 
\, + \,
{\cal{O}}\bigg(\Big(\frac{\alpha}{\pi}\Big)^3\bigg)
\,.
\label{Piloops}
\end{equation}
In the kinematic domain where $\alpha\ll|\beta|\ll1$, the expansion in
terms of the number of loops is still an adequate approximation, and
we are allowed to expand the coefficients in eq.~(\ref{Piloops}) for
small velocities,
\begin{eqnarray}
\Big(\frac{\alpha}{\pi}\Big)\,\Pi^{(1)}(q^2) 
& \stackrel{|\beta|\to 0}{=} &
\alpha\,\bigg[\, \frac{8}{9\,\pi } + \frac{i}{2}\,\beta \,\bigg]
 + {\cal{O}}(\alpha\,\beta^2)
\,,
\label{Pi1loopexpanded}
\\ [3mm]
\Big(\frac{\alpha}{\pi}\Big)^2\,\Pi^{(2)}(q^2) 
& \stackrel{|\beta|\to 0}{=} &
{{\alpha}^2}\,\bigg[\, \frac{1}{4\,{{\pi }^2}}\,
     \left( 3 - \frac{21}{2}\,\zeta_3 \right)  + \frac{11}{32} - 
    \frac{3}{4}\,\ln 2 - \frac{1}{2}\,\ln(-i\,\beta ) \,\bigg]
 + {\cal{O}}(\alpha^2\,\beta) 
\,,
\label{Pi2loopexpanded}
\end{eqnarray}
where $\beta=\sqrt{1-4M^2/(q^2+i\epsilon)}$ and 
$\zeta_3=1.202056903\ldots$.
In eq.~(\ref{Pi1loopexpanded}) the
${\cal{O}}(\alpha\beta)$ contribution is also displayed, allowing for a
check of the normalization during the matching procedure.
Whereas the one-loop contribution, eq.~(\ref{Pi1loopexpanded}), can be
evaluated for $\beta\to 0$, indicating that the ${\cal{O}}(\alpha)$
contribution of the vacuum polarization function is of pure 
short-distance origin, the two-loop expression, eq.~(\ref{Pi2loopexpanded}),
diverges logarithmically  for vanishing $\beta$. This shows that beyond 
${\cal{O}}(\alpha)$ accuracy conventional
perturbation theory is inadequate for
$|\beta|\lsim\alpha$ due to long-distance effects which cannot be
calculated in terms of a finite number of loop diagrams. These 
long-distance effects can, on the other hand, be described adequately by
the vacuum polarization function calculated in the framework of
quantum mechanics, which {\it per constructionem} is valid in the
non-relativistic limit. However, the vacuum polarization function as
defined in eq.~(\ref{PGunrenormalized}) gives a divergent result,\footnote{
In eq.~(\ref{PGunrenormalizeddiv}) we can identify 
$\beta=\sqrt{1-4 M^2/(q^2+i\epsilon)}$ with $\sqrt{(E+i\epsilon)/M}$
because we are interested in ${\cal{O}}(\alpha^2)$ accuracy only.}
\begin{eqnarray}
\Pi_{Thr}^{0,{\cal{O}}(\alpha^2)}(q^2) & = &
\frac{2\,\alpha\,\pi}{M^2}\,\lim_{r\to 0} G^0_E(0,r)
\nonumber\\[2mm] & = &
\frac{2\,\alpha\,\pi}{M^2}\,\lim_{r\to 0} \bigg[\,
-i\,\frac{M^2\,\beta}{2\,\pi}\,e^{i M \beta r}\,\int_0^\infty
  e^{2 i M \beta r t}\,\bigg(\frac{1+t}{t}\bigg)^{i \frac{\alpha}{2
\beta}}\,
{\rm d}t
\,\bigg]
\nonumber \\[2mm] & = &
\alpha\,\bigg[\, \frac{1}{2\,M\,r} + \frac{i}{2}\,\beta \,\bigg]  + 
  {{\alpha}^2}\,\bigg[\, \frac{1}{2} - \gamma - 
     \frac{1}{2}\,\ln(-2\,i\,M\,r\,\beta) - 
     \frac{1}{2}\,{\Psi}\Big(1 - i\,\frac{\alpha}{2\,\beta}\Big) \,\bigg] 
\,,
\label{PGunrenormalizeddiv}
\end{eqnarray}
where $\gamma$ is the Euler constant and $\Psi$ represents the digamma
function,
\begin{eqnarray}
\gamma & = & \lim_{n\to\infty}\,\bigg[\,
-\ln n \, + \, \sum_{i=1}^{n}\frac{1}{i}
\,\bigg] 
\, = \, 0.5772156649\ldots
\,,
\nonumber \\[3mm]
\Psi(z) & = & \frac{d}{d z}\,\ln\Gamma(z)
\,. 
\nonumber
\end{eqnarray}
The divergences in eq.~(\ref{PGunrenormalizeddiv}) can be easily
understood if non-relativistic quantum mechanics is considered as a
low-energy effective theory which is capable of describing
long-distance physics close to the threshold (characterized by momenta
below the scale of the electron mass) but does not know {\it per se}
any short-distance effects coming from momenta beyond the scale of the
electron mass. This lack of information is indicated in
eq.~(\ref{PGunrenormalizeddiv}) by short distance (UV) divergences and
has to be cured by matching non-relativistic quantum mechanics to
QED. The result of this matching procedure is called
``non-relativistic quantum electrodynamics'' (NRQED)~\cite{Caswell1}.
In this light we have to regard relations~(\ref{PGunrenormalized}) and
(\ref{PGunrenormalizeddiv}) as unrenormalized, which we have indicated
by using the superscripts $0$.
\par
In the common approach the Lagrangian of NRQED is obtained by
introducing higher dimensional operators in accordance with the
underlying symmetries of the theory and by matching them to
predictions within conventional perturbation theory in QED in the
kinematical regime $\alpha\ll|\beta|\ll 1$ where both NRQED and QED
are valid and must give the same results\footnote{
For $\alpha\ll|\beta|\ll 1$ conventional perturbation theory in QED is valid
because $\alpha$ represents the smallest parameter, whereas NRQED is
valid because $|\beta|$ is much smaller than the speed of light.
}.
In general this leads to divergent renormalization constants
multiplying the NRQED operators which then cancel the divergences in
eq.~(\ref{PGunrenormalizeddiv}) and add the correct finite 
short-distance contributions. In our case, however, the explicit
determination of the renormalization constants is not necessary
because we can match the vacuum polarization function obtained from
unrenormalized NRQED directly to the one- and two-loop expressions
from QED, eqs.~(\ref{Pi1loopexpanded}) and (\ref{Pi2loopexpanded}). 
This ``direct matching'' method has the advantage that the
regularization of the UV divergences in
eq.~(\ref{PGunrenormalizeddiv}) can be performed in a quite sloppy
way, but has the disadvantage that it is of no value to determine
other quantities than the vacuum polarization function itself.
To arrive at the vacuum polarization function in the threshold region
we just have to replace the $\beta$-independent and divergent
contributions of the ${\cal{O}}(\alpha)$ and ${\cal{O}}(\alpha^2)$
coefficients of expression~(\ref{PGunrenormalizeddiv}) in an expansion
for small $\alpha$
\begin{eqnarray}
\Pi_{Thr}^{0,{\cal{O}}(\alpha^2)}(q^2) & = &
\alpha\,\bigg[\, \frac{1}{2\,M\,r} + \frac{i}{2}\,\beta \,\bigg] 
+ {{\alpha}^2}\,
   \bigg[\, \frac{1}{2} - \frac{1}{2}\,{\gamma} - 
     \frac{1}{2}\,\ln (-2\,i\,M\,r\,\beta) \,\bigg] \,\nonumber\,
   \\ [2mm]
 & & \mbox{} + A(\alpha,\beta) 
\,,
\label{PGunrenormalizeddiv2}
\\ [4mm]
A(\alpha,\beta) & \equiv & 
-\frac{\alpha^2}{2}\,\bigg[\,
\gamma + \Psi\Big(1-i\frac{\alpha}{2\,\beta}\Big)
\,\bigg]
\label{Adefinition}
\end{eqnarray}
by the corresponding $\beta$-independent and finite contributions
in eqs.~(\ref{Pi1loopexpanded}), (\ref{Pi2loopexpanded}). The
correct normalization between the contributions coming from the loop 
calculations and from non-relativistic quantum mechanics can be checked
explicitly by observing that the coefficients in front of the 
$\beta$-dependent ${\cal{O}}(\alpha)$ and ${\cal{O}}(\alpha^2)$ 
contributions in eqs.~(\ref{Pi1loopexpanded}),
(\ref{Pi2loopexpanded}) and
(\ref{PGunrenormalizeddiv2}) are identical.
At this point we would like to note
that the function $A$ contains only
contributions of order $\alpha^3$ and higher for an expansion in small
$\alpha$. We shall return to this point later. 
The final result for the vacuum polarization function valid to
${\cal{O}}(\alpha^2)$ accuracy in the threshold region then reads
\begin{eqnarray}
\Pi_{Thr}^{{\cal{O}}(\alpha^2)}(q^2) & = &
\alpha\,\bigg[\, \frac{8}{9\,\pi } + \frac{i}{2}\,\beta \,\bigg] \,\nonumber\,
   \\ [2mm]
 & & \mbox{} + {{\alpha}^2}\,
   \bigg[\, \frac{1}{4\,{{\pi }^2}}\,
      \left( 3 - {{21}\over 2}\,\zeta_3 \right)  + \frac{11}{32} - 
     \frac{3}{4}\,\ln 2 - \frac{1}{2}\,\ln(- i\,\beta ) \,\bigg]
    \,\nonumber\,\\ [2mm]
 & & \mbox{} + A(\alpha,\beta)
\,.
\label{Pithreshfinal}
\end{eqnarray}
It is an interesting fact that this result can be obtained directly
from the one- and two-loop results, eqs.~(\ref{Pi1loopexpanded}) and
(\ref{Pi2loopexpanded}), by the replacement
\begin{eqnarray}
\ln(-i\,\beta) \, \longrightarrow \,
H(\alpha,\beta) &  \equiv & 
\gamma + \ln(-i\,\beta) + \Psi\Big(1-i\frac{\alpha}{2\,\beta}\Big)
\nonumber \\[2mm] & = & 
\ln(-i\,\beta) - \frac{2}{\alpha^2}\,A(\alpha,\beta)
\,.
\label{Hdefinition}
\end{eqnarray}
As mentioned earlier, the function $A$ is of order $\alpha^3$ for
$\alpha\ll|\beta|$. In the language of Feynman diagrams $A$ arises
from diagrams with instantaneous Coulomb exchange of two and more
longitudinal photons (in Coulomb gauge) between the electron-positron
pair. 
However, if $|\beta|$ is smaller than $\alpha$ ($\beta$ being real),
then $A$ is of order $\alpha^2$,
\begin{equation}
A(\alpha,\beta) 
\,\stackrel{|\beta|\ll\alpha}{=}\,
\frac{\alpha^2}{2}\,\bigg[\,
\ln\Big(i\,\frac{2\,\beta}{\alpha}\Big) - \gamma
\,\bigg] \, + \, {\cal{O}}(\alpha\,\beta)
\,.
\label{Asmallb}
\end{equation}
At this point it is illustrative to examine the limits
$\alpha\ll\beta$ and $\beta\ll\alpha$ for the function $H$,
defined in eq.~(\ref{Hdefinition}), for real and positive
values of $\beta$:
\begin{eqnarray}
H(\alpha,\beta) 
& \stackrel{\alpha\ll\beta}{=} &
\ln(-i\,\beta) + {\cal{O}}\Big(\frac{\alpha}{\beta}\Big)
\,, \\[2mm] 
H(\alpha,\beta) 
& \stackrel{\beta\ll\alpha}{=} & 
\ln\Big(\frac{\alpha}{2}\Big) + \gamma -i\,\pi +
{\cal{O}}\Big(\frac{\beta}{\alpha}\Big)
\,.
\label{Hsmallbeta}
\end{eqnarray}
It is evident that the function $H$ interpolates between a
$\ln\beta$-behaviour in the region where conventional perturbation
theory is valid and a constant with a logarithm of $\alpha$ for
$\beta/\alpha\to 0$. This leads to a finite value for
$\Pi_{Thr}^{{\cal{O}}(\alpha^2)}$ at the threshold point.
As we will see in the next section,
$\Pi_{Thr}^{{\cal{O}}(\alpha^2)}$
has singularities at the positronium energy levels indicating that the
breakdown of conventional perturbation theory is directly related
to the formation of bound states of the virtual $e^+e^-$
pair~\cite{Braun1}.
\par
Based on result~(\ref{Pithreshfinal}) we are now able to define a
renormalized expression for the zero-distance Green function 
by inverting relation~(\ref{PGunrenormalized})
\begin{equation}
G^R_E(0,0) \, \equiv \,
\frac{M^2}{2\,\alpha\,\pi}\,
\Pi_{Thr}^{{\cal{O}}(\alpha^2)}(q^2)
\,,
\label{Greenfunctionrenormalized}
\end{equation}
As we will show later, this renormalized zero-distance Green function
can be used for the calculation of higher-order corrections in
time-independent perturbation theory.
\par
For completeness we also present the QED vacuum polarization function
with ${\cal{O}}(\alpha^2)$ accuracy for all energies,
\begin{eqnarray}
\lefteqn{ 
\Pi_{\mbox{\tiny QED}}^{{\cal{O}}(\alpha^2)}(q^2) 
 \, = \, 
\Big(\frac{\alpha}{\pi}\Big)\,\Pi^{(1)}(q^2) 
\, + \,
\Big(\frac{\alpha}{\pi}\Big)^2\,\Pi^{(2)}(q^2) 
\, + \,  A(\alpha,\beta)
}
\nonumber\\[2mm] & = &
\Big(\frac{\alpha}{\pi}\Big)\,\bigg\{\,
\frac{8 - 3\,{{\beta}^2}}{9} + \frac{\beta\,\left( 3 - {{\beta}^2} \right) }{6
    }\,\ln(-p)
\,\bigg\} 
\nonumber\\[2mm] & & +
\Big(\frac{\alpha}{\pi}\Big)^2\,\bigg\{\,
\frac{18 - 13\,{{\beta}^2}}{24} + 
  \frac{\beta\,\left( 5 - 3\,{{\beta}^2} \right) }{8}\,\ln(-p) - 
  \frac{\left( 1 - \beta \right) \,
     \left( 33 - 39\,\beta - 17\,{{\beta}^2} + 7\,{{\beta}^3} \right) }{96}\,
   {{\ln(-p)}^2}\,\nonumber\,\\ 
 & & \mbox{}\,\qquad + 
  \frac{\beta\,\left( -3 + {{\beta}^2} \right) }{3}\,
   \bigg[\, 2\,\ln(1 - p)\,\ln(-p) + \ln(-p)\,\ln(1 + p) + \mbox{Li}_2(-p) + 
     2\,\mbox{Li}_2(p) \,\bigg] \,\nonumber\,\\ 
 & & \mbox{}\,\qquad + 
  \frac{\left( 3 - {{\beta}^2} \right) \,\left( 1 + {{\beta}^2} \right) }{12
    }\,\bigg[\, 2\,\ln(1 - p)\,{{\ln(-p)}^2} + 
     {{\ln(-p)}^2}\,\ln(1 + p)\,\,\nonumber\,\\ 
 & & \mbox{}\,\qquad\,
      \qquad\, + 4\,\ln(-p)\,\mbox{Li}_2(-p) + 
     8\,\ln(-p)\,\mbox{Li}_2(p) - 6\,\mbox{Li}_3(-p) - 12\,\mbox{Li}_3(p) - 
     3\,\zeta_3 \,\bigg] 
\,\bigg\}
\nonumber \\[2mm] & &
+ A(\alpha,\beta)
\,,
\label{Piallenergies}
\end{eqnarray} 
where 
\[ p \, \equiv \, \frac{1-\beta}{1+\beta} 
\] 
and $\mbox{Li}_2$, $\mbox{Li}_3$ denote the di- and trilogarithms~\cite{Lewin1}.
The reader should note that $\Pi_{\mbox{\tiny QED}}^{{\cal{O}}(\alpha^2)}$ vanishes
at $q^2=0$ and is an analytic function in $q^2$ except at poles and
branch cuts, and satisfies the
dispersion relation
\begin{equation}
\Pi_{\mbox{\tiny QED}}^{{\cal{O}}(\alpha^2)}(q^2) \, = \,
\frac{q^2}{\pi}\,\int\limits_{-\infty}^\infty
\frac{{\rm d}{q^\prime}^2}{{q^\prime}^2}\,
\frac{1}{{q^\prime}^2-q^2-i\epsilon}\,
{\rm Im} \Pi_{\mbox{\tiny QED}}^{{\cal{O}}(\alpha^2)}({q^\prime}^2)
\,.
\end{equation}
The explicit form of ${\rm Im} \Pi_{\mbox{\tiny QED}}^{{\cal{O}}(\alpha^2)}$ 
in the threshold region will be presented in Section~\ref{SectionAnalysis}.
For the use and interpretation of formula~(\ref{Piallenergies}) see also 
Section~\ref{SectionThreshold}.
\par
\vspace{0.5cm}
\section{Examination of the Vacuum Polarization Function \\
in the Threshold Region}
\label{SectionAnalysis}
In this section we analyse the properties of the vacuum
polarization function in the threshold region
above and below the threshold point $q^2=4M^2$. 
Compared to an older work on the same subject~\cite{Braun1} we are not
so much interested in general properties of perturbation theory in the
presence of bound state formation, but in the explicit form and
behaviour of $\Pi_{Thr}^{{\cal{O}}(\alpha^2)}$. In particular, we
focus on the size of the ${\cal{O}}(\alpha^2)$ contributions.
We also would like to mention that the vacuum polarization function
has been studied in a similar way in~\cite{Barbieri1}. In the latter publication, 
however, a different definition for the vacuum polarization is employed,
only the positronium ground state energy is considered and a 
contribution in the one-loop result is missing. 
We will come back to this point later.
Comparing the methods used in~\cite{Barbieri1} with the 
effective field theoretical
approach employed in this work makes the elegance of the latter technique 
obvious.
\par
We start in the kinematic region above threshold where
$\alpha\ll\beta\ll 1$. Here, as mentioned in the previous
section, the one- and two-loop results, 
eqs.~(\ref{Piloops})-(\ref{Pi2loopexpanded}), are reliable. This is
consistent with the fact that the function $A$ contains only
contributions of order $\alpha^3$ and higher,
\begin{eqnarray}
A(\alpha,\beta) &
\stackrel{\alpha\ll\beta}{=} &
\frac{\alpha^2}{2}\,\sum\limits_{n=2}^{\infty}\,\zeta_n\,
    \Big(\,i\,\frac{\alpha}{2\,\beta}\,\Big)^{n-1}
\nonumber\\[2mm] & = & 
\alpha^3\,\bigg[\,\frac{i}{24}\frac{\pi^2}{\beta}\,\bigg] - 
\alpha^4\,\bigg[\,\frac{\zeta_3}{8\,\beta^2}\,\bigg] + 
{\cal{O}}\Big(\frac{\alpha^5}{\beta^3}\Big)
\,.
\label{Asmalla}
\end{eqnarray}
Thus, for practical applications the contributions of function $A$ can
be neglected. (See also the discussion in
Section~\ref{SectionThreshold}.)
One might think that for $\alpha\approx\beta$ the one- and two-loop
expressions should still represent an appropriate ${\cal{O}}(\alpha^2)$ 
prediction, because the radius of convergence of the series on the
r.h.s. of eq.~(\ref{Asmalla}) is $|\beta|=\alpha/2$~\cite{Abramowitz1}. 
\begin{figure}[ht]
 \begin{center}
 \leavevmode
\epsfxsize=5cm
\epsffile[220 420 420 550]{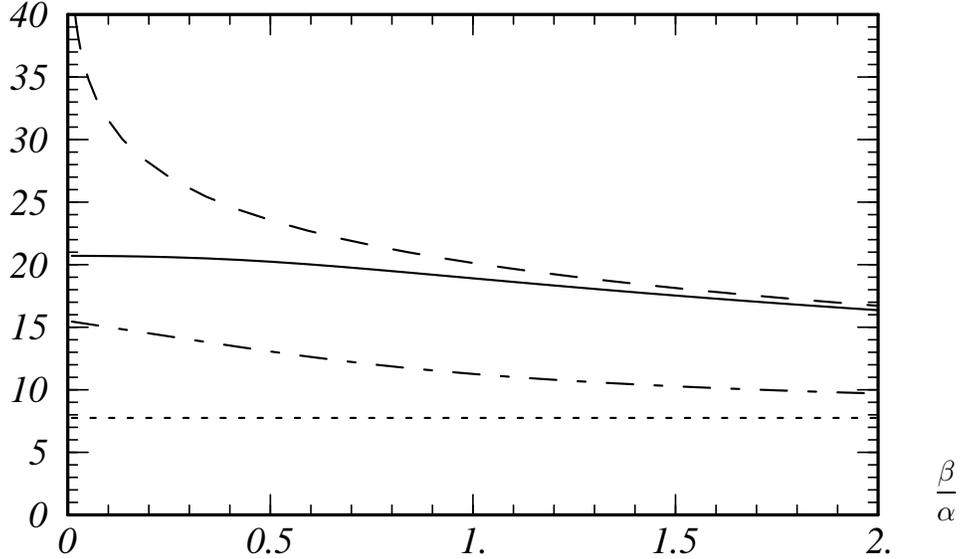}
\vskip 30mm  $\displaystyle{\mbox{\hspace{12.5cm}}\bf\frac{\beta}{\alpha}}$
\vskip 5mm
 \caption{\label{FigPi2} 
 The ${\cal{O}}(\alpha^2)$ corrections to the vacuum polarization function
 in the threshold region with and without the contributions contained in
 function $A$, eq.(\ref{Adefinition}), in the kinematic region $0<\beta<2\alpha$ 
 above the threshold.
 The solid line denotes $\mbox{Re}[\Pi^{(2)}+\pi^2/\alpha^2\,A]$, 
 the dashed line $\mbox{Re} \Pi^{(2)}$, 
 the dashed-dotted line $\mbox{Im}[\Pi^{(2)}+\pi^2/\alpha^2\,A]$
 and the dotted line $\mbox{Im} \Pi^{(2)}$.
 The value of the fine structure constant is taken as $\alpha=1/137$.
 $\Pi^{(2)}$ represents the two-loop contribution to the vacuum polarization
 function and is displayed for in eq.~(\ref{Pi2loopexpanded}).}
 \end{center}
\end{figure}
However, as
illustrated in Fig.~\ref{FigPi2}, for $\alpha\approx\beta$ the
contributions coming from function $A$ are already of order
$\alpha^2/\pi^2$ and thus have to be included if ${\cal{O}}(\alpha^2)$
accuracy is intended. For even smaller velocities, of course,  the
contributions from $A$ are essential because they cancel the divergent
$\ln\beta$ term from the two-loop expression $\Pi^{(2)}$,
see eqs.~(\ref{Pi2loopexpanded}) and (\ref{Asmallb}). Therefore the value of 
$\Pi_{Thr}^{{\cal{O}}(\alpha^2)}$ at the threshold point is finite and
reads\footnote{
The plus sign in the argument of $\Pi_{Thr}^{{\cal{O}}(\alpha^2)}$
indicates that the expression on the
r.h.s. of eq.~(\ref{Pithreshthresh}) represents only a right-sided limit
on the real $q^2$-axis.
} ($\alpha=1/137$)\\
\begin{eqnarray}
\lefteqn{
\Pi_{Thr}^{{\cal{O}}(\alpha^2)}(q^2\to 4M^2\,+) 
}\nonumber \\ [2mm]  & = &
\Big(\frac{\alpha}{\pi }\Big)\,\frac{8}{9} + 
  {{\alpha}^2}\,\bigg[\, -\frac{1}{2}\,\ln\alpha  - 
     \frac{1}{2}\,{\gamma} + 
     \frac{1}{4\,{{\pi }^2}}\,\left( 3 - \frac{21}{2}\,\zeta_3 \right)  + 
     \frac{11}{32} - \frac{1}{4}\,\ln 2 + i\,\frac{\pi }{2} \,\bigg] 
\nonumber \\[2mm] & = &
0.89\,\Big(\frac{\alpha}{\pi }\Big) + 
\Big(\,-0.36 - \frac{1}{2}\,\ln\alpha + i\,\frac{\pi}{2}\,\Big)\,\alpha^2
\, = \,
0.89\,\Big(\frac{\alpha}{\pi }\Big) + 
\Big(\,2.10 + i\,\frac{\pi}{2}\,\Big)\,\alpha^2
\,.
\label{Pithreshthresh}
\end{eqnarray}
It is evident from eq.~(\ref{Pithreshthresh}) and Fig.~\ref{FigPi2}
that the size of the ${\cal{O}}(\alpha^2)$ corrections in the threshold
region is of order $\alpha^2$ rather than $\alpha^2/\pi^2$, whereas the
${\cal{O}}(\alpha)$ contribution is of order $\alpha/\pi$. This
can be understood from the fact that the ${\cal{O}}(\alpha)$
contribution in $\Pi_{Thr}^{{\cal{O}}(\alpha^2)}$ comes entirely
from the one-loop result $\Pi^{(1)}$, eq.~(\ref{Pi1loopexpanded}), and
therefore originates from momenta beyond the scale of the electron
mass. High momenta contributions are expected to be of order
$\alpha/\pi$ if no ``large logarithms'' occur\footnote{
For comparison the reader might consider the well-known one- and two-loop
contributions to the anomalous magnetic moment of the 
electron~\cite{Schwinger1,Petermann1},
$g_e-2 =
(\frac{\alpha}{\pi})-0.66\,(\frac{\alpha}{\pi})^2+
{\cal{O}}((\frac{\alpha}{\pi})^3)$.
Here, long-distance effects from the $e^+e^-$ threshold play
no role. Therefore $g_e-2$ can be regarded as a typical
short-distance quantity with no ``large logarithms''.
}.
The large ${\cal{O}}(\alpha^2)$ contributions to
$\Pi_{Thr}^{{\cal{O}}(\alpha^2)}$, on the other hand, 
arise from the interplay of
the logarithm of the velocity in $\Pi^{(2)}$,
eq.~(\ref{Pi2loopexpanded}), and the contributions from the
instantaneous Coulomb-exchange of two and more longitudinal photons
between the virtual electron-positron pair. For small velocities the 
latter effects
generate a logarithm of the velocity  with an opposite sign, which
cancels the logarithm in  $\Pi^{(2)}$. We therefore
conclude that the large ${\cal{O}}(\alpha^2)$ contributions are of
long-distance origin. This is particularly obvious for the $\ln\alpha$
term which could never be generated at short distances.
\par
The situation for $\alpha\ll|\beta|\ll 1$ below threshold is similar to
the one above threshold. Here, the one- and two-loop contributions from
conventional perturbation theory,
eqs.~(\ref{Piloops})-(\ref{Pi2loopexpanded}), provide a viable
prediction, because the contributions from the function $A$ are of
order $\alpha^3$ and higher. 
They are beyond the intended accuracy and can be
neglected. (See also the discussion in
Section~\ref{SectionThreshold}.)
On the other hand, it is obvious that the one- and two-loop results
are not sufficient for energies close to the positronium bound state
energies, 
\begin{equation}
\beta \, = \, i\,\frac{\alpha}{2\,n} 
\,\, \Longleftrightarrow \,\,
E \, = \, -\frac{M\,\alpha^2}{4\,n^2} \,,
\qquad\qquad (n=1,2,3,\ldots)
\,,
\end{equation}
because the vacuum polarization function is expected to have poles at
those energy values. Therefore the full expression for
$\Pi_{Thr}^{{\cal{O}}(\alpha^2)}$, eq.~(\ref{Pithreshfinal}), must be
employed. It is straightforward to check that
$\Pi_{Thr}^{{\cal{O}}(\alpha^2)}$ indeed has poles at the positronium
energy levels~\cite{Braun1} leading to the following Laurent expansion at
the bound state energies $E_n= -\frac{M\,\alpha^2}{4\,n^2}$,
($n=1,2,3,\ldots$),
\begin{equation}
\lim\limits_{E\to E_n}\,
\Pi_{Thr}^{{\cal{O}}(\alpha^2)}(q^2) \, = \,
\frac{M\,\alpha^4}{4\,n^3}\,\frac{1}{E_n-E-i\epsilon} + 
\Big(\frac{\alpha}{\pi}\Big)\,\frac{8}{9} +
\alpha^2\,\bigg[\,\,a_n\,\bigg]
+ {\cal{O}}(E_n-E)
\,,
\label{PiLaurent}
\end{equation}
where
\begin{equation}
a_n \, \equiv \,
-\frac{1}{2}\,\ln\alpha + 
\frac{1}{2}\,\bigg[\,\frac{1}{n}+\ln n - 
    \sum\limits_{i=1}^{n-1}\frac{1}{i}\,\bigg] +
\frac{1}{4\,\pi^2}\,\bigg(\,3-\frac{21}{2}\,\zeta_3\,\bigg) +
\frac{11}{32} - \frac{1}{4}\,\ln 2
\,.
\label{andefinition}
\end{equation}
For completeness we also present the corresponding Laurent expansion
for the renormalized zero-distance Green function based on
definition~(\ref{Greenfunctionrenormalized}),
\begin{equation}
\lim\limits_{E\to E_n}\,
G^R_{E}(0,0) \, = \, 
\frac{\,\,\left|\Psi_n(0)\right|^2}{E_n-E-i\epsilon} + 
\frac{4}{9}\,\frac{M^2}{\pi^2} + 
\frac{M^2\,\alpha}{2\,\pi}\,\bigg[\,a_n\,\bigg]
+ {\cal{O}}(E_n-E)
\,.
\label{Greenfunctionresidues}
\end{equation}
As expected, the residues at the bound state energies are equal to the
moduli squared of the normalized $l=0$ Coulomb wave functions at the origin, 
\begin{equation}
\left|\Psi_n(0)\right|^2 \, = \, \frac{M^3\,\alpha^3}{8\,\pi\,n^3}
\,.
\label{wavefunctionsquared}
\end{equation}
In eqs.~(\ref{PiLaurent})-(\ref{Greenfunctionresidues}) we have also
displayed the constant terms of the Laurent expansion. These constants
are relevant for higher-order corrections to the positronium energy
levels and to the wave functions at the origin. The size of the
${\cal{O}}(\alpha^2)$ corrections in these constant terms is (similar
to the ${\cal{O}}(\alpha^2)$ contributions above threshold) of order
$\alpha^2$ rather than $\alpha^2/\pi^2$ indicating again the
long-distance character of the ${\cal{O}}(\alpha^2)$ corrections. 
\begin{table}[htb]
\vskip 7mm
\begin{center}
\begin{tabular}{|c||c|c|c|c|c|c|} \hline
$n$ & $1$ &  $2$ & $3$ & $4$ & $5$ & $\infty$ \\ \hline\hline
\parbox{.4cm}{\vskip 2mm $a_n$ \vskip 2mm} & 
    $2.89$ & $2.48$ & $2.35$ & $2.29$ & $2.25$ & $2.10$ \\ \hline 
\end{tabular}
\caption{\label{Tablean} 
 The numerical value for
 the constants $a_n$ for the radial quantum numbers $n=1,2,3,4,5$ and for
 $n\to\infty$ with $\alpha=1/137$.
}
\end{center}
\vskip 3mm
\end{table}
In Table~\ref{Tablean} we
have displayed the numerical values of $a_n$ for the radial quantum
numbers $n=1,2,3,4,5$. It is an interesting fact that the $n\to\infty$
limit of $a_n$ exists
\begin{equation}
\lim\limits_{n\to\infty}\,\alpha^2\,\bigg[\,a_n\,\bigg] \, = \,
\alpha^2\,\bigg[\,
- \frac{1}{2}\,\ln\alpha - \frac{1}{2}\,{\gamma} + 
  \frac{1}{4\,{{\pi }^2}}\,\left( 3 - \frac{21}{2}\,\zeta_3 \right)  + 
  \frac{11}{32} - \frac{1}{4}\,\ln 2
\,\bigg]
\end{equation} 
and coincides with ${\cal{O}}(\alpha^2)$ contributions of $\mbox{Re}
\Pi_{Thr}^{{\cal{O}}(\alpha^2)}(q^2\to 4M^2 +)$,
eq.~(\ref{Pithreshthresh}). The numerical value for 
$\lim_{n\to\infty}\,a_n$ is presented in Table~\ref{Tablean}.
\par
To illustrate the importance of the constants $a_n$ in 
time-independent perturbation theory (TIPT), we recalculate 
the  ${\cal{O}}(\alpha^6)$ ``vacuum polarization'' contributions to
the ground state triplet-singlet 
hyperfine splitting (HFS) of the positronium, which were, to our
knowledge, considered for the first time in~\cite{Barbieri1}. The vacuum
polarization contributions to the HFS in the energy levels of the
positronium system arise from the effect that the bound triplet
(${}^3S_1$, $J^{PC}=1^{-\,-}$) $e^+e^-$ pair can annihilate into a
virtual photon for a time period of order $1/M$, whereas the singlet  
(${}^1S_0$, $J^{PC}=0^{-\,+}$) cannot. If the virtual photon energy is
approximated by $\sqrt{q^2}=2M$, this annihilation process leads to a
$\delta$-function kernel in the coordinate-space representation
(corresponding to a constant kernel in momentum space) with the form,
\begin{equation}
H_{Ann}(\vec{x}) \, = \,
\frac{2\,\alpha\,\pi}{M^2}\,\delta^{(3)}(\vec{x})
\,.
\label{AnnihilationKernel}
\end{equation}
This kernel can now be used in  TIPT. Taking into account that
$\Pi_{Thr}^{{\cal{O}}(\alpha^2)}$ contains ${\cal{O}}(\alpha)$ as well as 
${\cal{O}}(\alpha^2)$ contributions we have to apply second- and 
third-order TIPT to obtain all relevant ${\cal{O}}(\alpha^6)$ contributions to
the HFS. The formal result for the ${\cal{O}}(\alpha^6)$ energy shift for 
the triplet states with radial quantum numbers $n$ and with $l=0$ 
due to $H_{Ann}$ reads
\begin{eqnarray}
\delta E^{\alpha^6}_{Ann,n} & = &
\bigg\{\,
\sum\hspace{-5.5mm}\int\limits_{l\ne n}\,\,\,
\langle\,n\,| \, H_{Ann} \, \frac{|\,l\,\rangle\,\langle \,l\,|}{E_n-E_l} \,
              H_{Ann} \, |\,n\,\rangle
\nonumber\\[2mm] & & \,\, + \,
\sum\hspace{-6.3mm}\int\limits_{m\ne n}\,\,\,
\sum\hspace{-5.7mm}\int\limits_{k\ne n}\,\,\,
\langle\,n\,| \, H_{Ann} \, \frac{|\,m\,\rangle\,\langle \,m\,|}{E_n-E_m} \,
              \, H_{Ann} \, \frac{|\,k\,\rangle\,\langle \,k\,|}{E_n-E_k} \,
              H_{Ann} \, |\,n\,\rangle
\,\bigg\}_{{\cal{O}}(\alpha^6)}
\,,
\label{HFSformal}
\end{eqnarray}
where $|\,i\,\rangle$, $i=l,m,n,k$, represent normalized (bound state
and free scattering) eigenfunctions to the positronium Schr\"odinger
equation with the eigenvalues $E_i$. The symbol $\{\}_{{\cal{O}}(\alpha^6)}$
indicates that only ${\cal{O}}(\alpha^6)$ contributions are taken into account.
It is evident from the form of $H_{Ann}(\vec{x})$ that only the 
zero-distance Green function is relevant for $\delta E^{\alpha^6}_{Ann,n}$,
\begin{eqnarray}
\sum\hspace{-5.5mm}\int\limits_{l\ne n}\,\,\,
\langle\,0\,|\, \frac{|\,l\,\rangle\,\langle \,l\,|}{E_l-E_n}
\,|\,0\,\rangle
& = &
\sum\hspace{-5.5mm}\int\limits_{l\ne n}\,\,\,
\frac{\,\,|\Psi_l(0)|^2}{E_l-E_n}
\nonumber\\[2mm] & = &
\lim\limits_{E\to E_n}\,\bigg[\,
G_E^0(0,0) - \frac{\,\,|\Psi_n(0)|^2}{E_n-E-i\epsilon}
\,\bigg]
\,.
\label{Greenfunctionsubtracted}
\end{eqnarray}
However, relation~(\ref{Greenfunctionsubtracted}) still contains
divergences (see eq.(\ref{PGunrenormalizeddiv})). As we have pointed
out in Section~\ref{SectionCalculation}, these divergences indicate
that non-relativistic quantum mechanics is not capable to describe
physics if the relative distance of the electron-positron pair is
smaller than the inverse electron mass. Therefore, we have to replace
$G_E^0(0,0)$ in relation~(\ref{Greenfunctionsubtracted}) by its
renormalized version $G_E^R(0,0)$,
eq.~(\ref{Greenfunctionrenormalized}) (using eq.~(\ref{Pithreshfinal})), 
which describes short-distance
physics properly. The final expression for $\delta
E^{\alpha^6}_{Ann,n}$ then reads
\begin{eqnarray}
\lefteqn{
\delta E^{\alpha^6}_{Ann,n} \, = \,
|\Psi_n(0)|^2 \, 
\bigg\{\,
\bigg[\,\frac{2\,\alpha\,\pi}{M^2}\,\bigg]^2\,
\bigg(\,-\frac{M^2\,\alpha}{2\,\pi}\,a_n\,\bigg) + 
\bigg[\,\frac{2\,\alpha\,\pi}{M^2}\,\bigg]^3\,
\bigg(\,-\frac{4}{9}\,\frac{M^2}{\pi^2}\,\bigg)^2
\,\bigg\}
}
\nonumber\\[2mm] & = & 
\frac{M\,\alpha^6}{4\,n^3}\,\bigg\{\,
\frac{1}{2}\,\ln\alpha - 
\frac{1}{2}\,\bigg(\,\frac{1}{n} + \ln n - 
         \sum\limits_{i=1}^{n-1}\frac{1}{i}\,\bigg) +
\frac{1}{4\,\pi^2}\,\bigg(\,\frac{13}{81}+\frac{21}{2}\,\zeta_3\,\bigg)
-\frac{11}{32} + \frac{1}{4}\,\ln 2
\,\bigg\}
\,.
\label{HFSfinal}
\end{eqnarray}
Taking also into account that the virtual photon energy is smaller than
$2M$ by the amount of the binding energy
$E_n=-\frac{M\,\alpha^2}{4\,n^2}$ we have to replace $M^2$ 
in eq.~(\ref{AnnihilationKernel}) by $(M+E_n/2)^2$ and
therefore get another
${\cal{O}}(\alpha^6)$ contribution,
\begin{equation}
\delta E^{\alpha^6}_{Bind,n} \, = \,
\frac{\alpha^2}{4\,n^2}\,\delta E^{\alpha^4}_{Ann,n} \, = \,
\frac{M\,\alpha^6}{16\,n^5}
\,,
\label{HFSbinding}
\end{equation}
where $\delta E^{\alpha^4}_{Ann,n}$ represents the 
${\cal{O}}(\alpha^4)$ energy shift due to $H_{Ann}$. For the
ground state ($n=1$) the complete ${\cal{O}}(\alpha^6)$ vacuum
polarization contribution to the HFS then reads
\begin{equation}
\delta E^{\alpha^6}_{Ann,1} + \delta E^{\alpha^6}_{Bind,1}
\, = \,
\frac{M\,\alpha^6}{4}\,\bigg\{\,
\frac{1}{2}\,\ln\alpha + 
\frac{1}{4\,\pi^2}\,\bigg(\,\frac{13}{81}+\frac{21}{2}\,\zeta_3\,\bigg)
-\frac{19}{32} + \frac{1}{4}\,\ln 2
\,\bigg\}
\,.
\label{HFSgroundstate}
\end{equation}
Our result differs from the one presented 
in~\cite{Barbieri1,Barbieri2} by the
amount $M\alpha^6/8$. Half of the discrepancy
comes from the fact that in~\cite{Barbieri1} the binding energy contribution 
$\delta E^{\alpha^6}_{Bind,1}$ was not taken into account, whereas the
other half originates from a missing ${\cal{O}}(\alpha^2)$
contribution in the one-loop vacuum polarization\footnote{
This can be easily seen by comparing eq.~(33) of~\cite{Barbieri1} with
eq.~(\ref{Pi1loopexpanded}) in this work for
\[
q^2_{n=1}=(2\,M-\frac{M\alpha^2}{4})^2\,\,\Longleftrightarrow\,\,
\beta_{n=1}=i\,\frac{\alpha}{2} + {\cal{O}}(\alpha^3)
\,.
\]
This shows that at bound state energies the one-loop contribution to 
the vacuum polarization function also contains terms of order $\alpha^2$.
}.
\par
Before we turn to applications of our results in the context of QCD,
we do not want to leave unmentioned that the leading contributions to
the normalized cross section for production of a heavy
lepton-antilepton pair (with lepton mass $M$) in $e^+e^-$ collisions 
(via a virtual photon) in the threshold region can be recovered from
$\Pi_{Thr}^{{\cal{O}}(\alpha^2)}$ by means of the optical
theorem~\cite{Fadin1,Strassler1},
\begin{eqnarray}
R^{L^+L^-}_{Thr} & = & \frac{\sigma(e^-e^+\to\gamma^*\to
L^+L^-)}{\sigma_{pt}}
\, = \,
\frac{3}{\alpha}\,\mbox{Im}\Pi_{Thr}^{{\cal{O}}(\alpha^2)}(q^2)
\, = \,
\frac{6\,\pi}{M^2}\,\mbox{Im}\,G^R_E(0,0)
\nonumber\\[2mm] & = &
\frac{6\,\pi^2}{M^2}\,\sum\limits_{n=1}^{\infty}|\Psi_n(0)|^2\,
       \delta(E-E_n) \, + \,
\Theta(E)\,\frac{3}{2}\,\frac{\alpha\,\pi}{1-\exp(-\frac{\alpha\,\pi}{\beta})}
\,,
\label{OpticalTheorem}
\end{eqnarray}
where $\sigma_{pt}$ represents the point cross section and 
only final-state interactions are taken into account.
\par
\vspace{.5cm}
\section{Darwin Corrections in QCD}
\label{SectionQCD}
In the previous section we have shown that the size of the
${\cal{O}}(\alpha^2)$ corrections to the QED vacuum polarization 
function in the threshold
region is of order $\alpha^2$ rather than $\alpha^2/\pi^2$. Although
this fact is important for precision tests of QED\footnote{
As far as tests of QED in the $\tau^+\tau^-$ system in the 
threshold region are concerned the
present experiments do not even reach the ${\cal{O}}(\alpha)$ 
(next-to-leading order) accuracy level. This can be easily seen from 
the fact that the complete threshold region for the $\tau^+\tau^-$ system,
$|\beta|\lsim\alpha\,\Longleftrightarrow\,
|\sqrt{q^2}-2m_\tau|\lsim m_\tau\alpha^2 = 0.1~\mbox{MeV}$ still lies
within the limits on the tau mass itself, $m_\tau =
1777.00^{+0.30}_{-0.27}~\mbox{MeV}$~\cite{PDG}. Thus only experiments on
electron and muon systems can be regarded as precision tests of QED in
the threshold regime.
},
it does not lead to theoretical concerns about the convergence of
the perturbative series because of the smallness of the fine
structure constant $\alpha$ and because QED is not asymptotically
free.
\par
In the framework of QCD, however, the situation is completely
different: the coupling is much larger and even becomes of order one
for scales much lower than $1$~GeV. Therefore, the fact that the size of
the ${\cal{O}}(\alpha_s^2)$ (next-to-next-to-leading order) corrections in
the threshold region might be of order $\alpha_s^2$ rather than
$\alpha_s^2/\pi^2$ is an extremely important theoretical issue because
this would lead to corrections of order $1\%-25\%$ rather than
$0.1\%-2.5\%$ for $\alpha_s=0.1-0.5$. Here, two natural questions arise:
what scale should be used in the strong coupling, and
for which heavy quark-antiquark systems the ${\cal{O}}(\alpha_s^2)$ 
corrections represent contributions to the asymptotic perturbation 
series in the convergent regime. These questions will be addressed 
in the following section.
\par 
To be more specific we will calculate the ${\cal{O}}(C_F^2\alpha_s^2)$
Darwin corrections to the ($S$-wave, $l=0$) wave functions of a bound heavy
quark-antiquark pair at the origin and to the
heavy quark-antiquark pair production cross section
in $e^+e^-$ annihilation (via a
virtual photon) in the threshold region. 
A presentation of all ${\cal{O}}(C_F^2\alpha_s^2)$ corrections
including all kinematic and relativistic effects will given in a
subsequent publication.
The corresponding uncorrected
quantities are the well-known exact solutions to a pure Coulomb-like
non-relativistic quark-antiquark system described by a Schr\"odinger
equation with the QCD-potential $V_{QCD}(r)=-C_F\,\frac{\alpha_s}{r}$,
where $C_F=(N_c^2-1)/2N_c=4/3$.
\par
The Darwin interaction is generated in the non-relativistic expansion
of the Dirac equation. In the coordinate-space representation it is
proportional to a $\delta$-function and reads\footnote{
Compared to the Darwin interaction known from the hydrogen atom the
expression on the r.h.s. of eq.~(\ref{Darwindef}) is a factor of
two larger because both quark-antiquark-gluon vertices contribute.
}
\begin{equation}
H_{Dar}(\vec{x}) \, = \, \frac{C_F\,\alpha_s\,\pi}{M_Q^2}\,
\delta^{(3)}(\vec{x})
\,.
\label{Darwindef}
\end{equation}
A practical application for the corrections to the bound state wave
functions at the origin is the leptonic decay rate of the
$J/\Psi$ and the $\Upsilon(1S)$ and (maybe) of the first few excited
states of the $\Upsilon$ family, whereas the corrections to the cross
section would be relevant for $t\bar t$ production at NLC. We
explicitly mention those applications in this context because it is
believed that for them non-perturbative (in the sense ``not calculable
from first principles in QCD'') effects are either well under control
or even negligible~\cite{Yndurain1,Fadin1}. 
But, of course, these corrections can be
applied to other heavy quark-antiquark systems as well, 
at least in order to
check their size. At this point we want to emphasize that 
we do not intend to present a thorough phenomenological analysis
in this work. The primary aim is to use the 
${\cal{O}}(C_F^2\alpha_s^2)$ Darwin corrections to illustrate
the typical size of the complete (and  yet unknown) 
${\cal{O}}(\alpha_s^2)$ corrections for the $t\bar t$,
$b\bar b$ and $c\bar c$ systems. Their actual numerical value and even 
their sign cannot, of course, be predicted at the present stage.
\par
To keep our analysis transparent we ignore all ${\cal{O}}(\alpha_s)$
corrections, the effects from the running of the strong
coupling and also non-perturbative contributions like the gluon
condensate. The latter effects are well known and have been treated in
a large number of earlier publications. 
We further neglect the width
of the quarks and treat them as stable particles for the most part
in the following analysis. From the technical point of view 
the calculations of the ${\cal{O}}(C_F^2\alpha_s^2)$ Darwin
corrections are identical to the corresponding QED calculations, which
means that we use time-independent perturbation theory. However, we
have to take care about the correct implementation of the number of
colors, $N_c=3$, and the group theoretical factor $C_F$. In the
following the superscript ``QCD'' indicates that the corresponding
quantity is obtained from the QED expression by the replacement
$\alpha\to C_F\alpha_s$. 
It is then straightforward to determine the
${\cal{O}}(C_F^2\alpha_s^2)$ Darwin corrections to the modulus squared
of the $l=0$ bound state wave functions at the origin, 
($n=1,2,3,\ldots$),
\begin{eqnarray}
\delta|\Psi_n^{\mbox{\tiny QCD}}(0)|^2_{Dar} & = &
-2\,|\Psi_n^{\mbox{\tiny QCD}}(0)|\,\bigg\{\,
\frac{C_F\,\alpha_s\,\pi}{M_Q^2}\hspace{-3mm}
\lim\limits_{\mbox{}\quad E->E_n^{\mbox{\tiny QCD}}}\,
\bigg[\,G^{R,\mbox{\tiny QCD}}_E(0,0) - 
    \frac{\,\,|\Psi_n^{\mbox{\tiny QCD}}(0)|^2}{E_n^{\mbox{\tiny QCD}}-E-i\epsilon}
\,\bigg]
\,\bigg\}_{{\cal{O}}(C_F^2\alpha_s^2)}
\nonumber\\[2mm] & = &
-\,|\Psi_n^{\mbox{\tiny QCD}}(0)|^2\,C_F^2\,\alpha_s^2\,a_n^{\mbox{\tiny QCD}}
\,,
\label{Darwinwavefunctions}
\end{eqnarray}
where the symbol $\{\}_{{\cal{O}}(C_F^2\alpha_s^2)}$ indicates that
only ${\cal{O}}(C_F^2\alpha_s^2)$ corrections are taken into
account\footnote{
Eq.~(\ref{Darwinwavefunctions}) also generates 
${\cal{O}}(C_F\alpha_s)$ corrections which differ from the 
well-known ${\cal{O}}(C_F\alpha_s)$ corrections generated by the
$(1-4 C_F\alpha_s/\pi)$ correction factor~\cite{Karplus1}. Adding
up all the ${\cal{O}}(C_F\alpha_s)$ corrections and the corresponding
renormalization constants will of course yield the correct
result. The same remark holds for the result for the cross section
above threshold, eq.~(\ref{Darwincrosssection}).
}.
The calculation of the ${\cal{O}}(C_F^2\alpha_s^2)$ Darwin corrections to the
quark-antiquark cross section in the threshold region is more
involved. Here, we apply the optical theorem,
eq.~(\ref{OpticalTheorem}), to the corrections of the zero-distance
Green function themselves,
\begin{equation}
\delta G_{E, Dar}^{R,\mbox{\tiny QCD}}(0,0) \, = \,
-\,\frac{C_F\,\alpha_s\,\pi}{M_Q^2}\,\bigg[\,
G_E^{R,\mbox{\tiny QCD}}(0,0)
\,\bigg]^2
\,.
\label{DarwinGreenfunction}
\end{equation}
The  ${\cal{O}}(C_F^2\alpha_s^2)$ Darwin corrections to the
cross section above threshold then read
\begin{eqnarray}
\delta R^{Q\bar Q}_{Thr, Dar} & = &
N_c\,\frac{6\,\pi}{M_Q^2}\,\mbox{Im} G^{R,\mbox{\tiny QCD}}_E(0,0)\,
\bigg\{\,
-\,\frac{2\,C_F\,\alpha_s\,\pi}{M_Q^2}\,
 \mbox{Re}\, G^{R,\mbox{\tiny QCD}}_E(0,0)
\,\bigg\}_{{\cal{O}}(C_F^2\alpha_s^2, C_F\alpha_s\beta)}
\nonumber\\[2mm] & = &
R^{Q\bar Q}_{Thr}\,\bigg\{\,
- \mbox{Re} \Pi_{Thr}^{{\cal{O}}(C_F^2\alpha_s^2),\mbox{\tiny QCD}}(q^2)
\,\bigg\}_{{\cal{O}}(C_F^2\alpha_s^2, C_F\alpha_s\beta)}
\,,
\label{Darwincrosssection}
\end{eqnarray} 
where $R^{Q\bar Q}_{Thr}$ represents the ``Sommerfeld factor''
(sometimes also called the ``Fermi factor'')
\begin{eqnarray}
\lefteqn{
R^{Q\bar Q}_{Thr} \,  = \, N_c\,
\frac{6\,\pi}{M_Q^2}\,\mbox{Im} G^{R,\mbox{\tiny QCD}}_E(0,0)
}
\nonumber\\[2mm] & = & 
 N_c\,\frac{3}{2}\,
\frac{C_F\,\alpha_s\,\pi}{1-\exp(-\frac{C_F\,\alpha_s\,\pi}{\beta})}
\, = \,
N_c\,\frac{3}{2}\,\beta\,
\exp\Big(\frac{C_F\,\alpha_s\,\pi}{2\,\beta}\Big)\,
\Gamma\Big(1+i\,\frac{C_F\,\alpha_s}{2\,\beta}\Big)\,
\Gamma\Big(1-i\,\frac{C_F\,\alpha_s}{2\,\beta}\Big)
\,.
\label{Sommerfeldfactor}
\end{eqnarray}
Below threshold we have to determine the corrections to the residues 
of $G_E^{R,\mbox{\tiny QCD}}(0,0)$ at the bound state energies, where
as shown above the corresponding bound state poles have to be
subtracted. This calculation is straightforward and leads to the
corrections to the $l=0$ bound state wave functions at the origin 
presented in eq.~(\ref{Darwinwavefunctions}).
It is an interesting fact that eq.~(\ref{Darwincrosssection}) allows
for the calculation of the shifts of the $Q\bar Q$ bound state
energies due to the Darwin interaction. To show this we rewrite the sum
of the Sommerfeld factor, eq.~(\ref{Sommerfeldfactor}), and the 
contribution involving the digamma function of the 
${\cal{O}}(C_F^2\alpha_s^2)$ Darwin corrections above threshold, see
eqs.~(\ref{Adefinition}), (\ref{Pithreshfinal}) and 
(\ref{Darwincrosssection}), as
\begin{eqnarray}
\lefteqn{
R^{Q\bar Q}_{Thr}\,\bigg\{\, 1\, + \, 
\frac{C_F^2\,\alpha_s^2}{4}\,
\bigg[\,
\Psi\Big(1+i\,\frac{C_F\,\alpha_s}{2\,\beta}\Big) +
\Psi\Big(1-i\,\frac{C_F\,\alpha_s}{2\,\beta}\Big)
\,\bigg]
\,\bigg\}
}
\nonumber\\[2mm] & \longrightarrow & 
N_c\,\frac{3}{2}\,\beta\,
\exp\Big(\frac{C_F\,\alpha_s\,\pi}{2\,\beta}\Big)\,
\Gamma\Big(\frac{C_F^2\,\alpha_s^2}{4}+1+
      i\,\frac{C_F\,\alpha_s}{2\,\beta}\Big)\,
\Gamma\Big(\frac{C_F^2\,\alpha_s^2}{4}+1-
      i\,\frac{C_F\,\alpha_s}{2\,\beta}\Big)
\,.
\label{Energyshiftindirect}
\end{eqnarray}
It can be easily checked that the function
$\Gamma(\frac{C_F^2\,\alpha_s^2}{4}+1-i\,\frac{C_F\,\alpha_s}{2\,\beta})$
develops poles at the energies\footnote{
It should be noted that the $(1-4 C_F\alpha_s/\pi)$ correction
factor of the cross section is irrelevant for shifts of the bound
state energies because the former represents a global multiplicative
short-distance factor.
}
\begin{equation}
\tilde{E}_n^{\mbox{\tiny QCD}} \, \equiv \,
E_n^{\mbox{\tiny QCD}} + \delta E_{n,Dar}^{\mbox{\tiny QCD}}
\,,\qquad\qquad
(n=1,2,3,\ldots)\,,
\end{equation}
where the $\delta E_{n,Dar}^{\mbox{\tiny QCD}}$ represent the energy
shift of the $l=0$ Coulomb energy levels with the radial quantum number $n$
generated by the Darwin interaction,
\begin{eqnarray}
\delta E_{n,Dar}^{\mbox{\tiny QCD}} & = & 
\langle\,n^{\mbox{\tiny QCD}}\,|\,H_{Dar}\,
|\,n^{\mbox{\tiny QCD}}\,\rangle \, = \,
|\Psi_n^{\mbox{\tiny QCD}}(0)|^2\,\frac{C_F\,\alpha_s\,\pi}{M_Q^2}
\nonumber \\[2mm] & = &
\frac{M_Q\,C_F^4\,\alpha_s^4}{8\,n^3}
\,.
\label{Energyshiftdirect}
\end{eqnarray} 
At this point we also want to emphasize that the $\ln(C_F\,\alpha_s)$
and digamma contributions
occurring in eqs.~(\ref{Darwinwavefunctions}) and
(\ref{Darwincrosssection}) are not related to the running of the strong
coupling and therefore cannot be resummed by any known type of
renormalization group equation in the sense of a leading logarithmic 
resummation. These logarithmic terms arise because two scales are relevant
in the threshold region, the heavy quark mass $M_Q$ and 
the relative momentum of the quark-antiquark pair $\propto
C_F\,\alpha_s\,M_Q$~\cite{Labelle1}. The $\ln(C_F\,\alpha_s)$
contributions induced by the running of the strong coupling have been
discussed in~\cite{Voloshin2,Yndurain1}.
\par
Before we turn to the discussion on the size of the
${\cal{O}}(C_F^2\alpha_s^2)$ Darwin corrections we have to address
the question of which scale one should use in the strong
coupling. Strictly speaking, a final answer to this problem would
require an ${\cal{O}}(\alpha_s^3)$ analysis, which is beyond the scope
of this work. However, one can find simple arguments that the scale in
the strong couplings of 
expressions~(\ref{Darwinwavefunctions}) and (\ref{Darwincrosssection})
should be of the order $C_F\alpha_s M_Q$, which will be called ``the
soft scale'' in the remainder of this work. We would like to remind
the reader that the scale of the strong coupling in the unperturbed
(pure Coulomb) quantities $|\Psi_n^{\mbox{\tiny QCD}}(0)|^2$ and 
$R^{Q\bar Q}_{Thr}$ is of the order of the soft scale. This is
obvious for the wave functions of the ground state and the first few
excited states and for the cross section in the kinematic region
$\beta\approx C_F\alpha_s$
because they describe bound quark-antiquark pairs with
relative momentum of order $C_F\,\alpha_s\,M_Q$. But it is also true
for highly excited states ($n\gg 1$) and the cross section right at the
threshold due to ``saturation'' effects~\cite{Voloshin2,Yndurain1}, 
{\it i.e.} the scale of
the strong coupling is of order  $C_F\alpha_s M_Q$ although the
kinematical relative momentum of the quark pair vanishes\footnote{
In~\cite{Voloshin2,Yndurain1} a proof for saturation is only given
for the cross section above the threshold point. An analogous proof
for highly excited stated or the cross section slightly below the
threshold point does, to our knowledge, not exist in literature.
Such a proof is, however, much more more difficult due to the
breakdown of time-independent perturbation theory for the
logarithmic kernel $\delta V(r)\sim\ln(r)/r$ for
high radial excitations (see {\it e.g.}~\cite{Yndurain1}).
Nevertheless we find it plausible that saturation also takes place
slightly below the threshold point because the cross section at
the threshold point $q^2=4M_Q^2$ should be well defined.
}.
To understand that the scale of the ${\cal{O}}(C_F^2\alpha_s^2)$
Darwin corrections should also be of order of the soft scale, let us 
have a closer look at the origin of the strong couplings governing these 
corrections: one power of $\alpha_s$ comes from the Darwin
interaction, $H_{Dar}$, and the other power of $\alpha_s$ (including
the $\ln(C_F\alpha_s)$ terms) originates
from the ${\cal{O}}(\alpha_s^2)$ contribution of 
the vacuum polarization function 
$\Pi_{Thr}^{{\cal{O}}(C_F^2\alpha_s^2),\mbox{\tiny QCD}}$. As
mentioned in the previous section, the latter contribution is of
long-distance origin and therefore governed by the soft scale. In
contrast to the pure Coulomb interaction, $1/\vec{p}^2$, the Darwin
interaction is a constant in momentum space and consequently sensitive
to both low and high momenta. But based on our previous observations of
the domination of long-distance effects, we can assume that the scale
of the strong coupling in the Darwin interaction should also be the
soft scale rather than the heavy quark mass. The size
of the strong coupling governing the ${\cal{O}}(C_F^2\alpha_s^2)$
Darwin corrections of eqs.~(\ref{Darwinwavefunctions}) and
(\ref{Darwincrosssection}) can therefore be estimated via the
self-consistency equation
\begin{equation}
\alpha_s = \alpha_s(C_F\,\alpha_s\,M_Q)
\,.
\label{Selfconsistency}
\end{equation}
which leads to $\alpha_s=0.13-0.16$, $0.25-0.38$ and $0.34-0.59$ for
the top, bottom and charm quark systems, respectively. The latter ranges
are obtained by using the $\overline{\mbox{MS}}$ definition for the
strong coupling, the one-loop QCD beta-function and
$\alpha_s(M_z=91.187\,\mbox{GeV})=0.125$ and by taking twice and half the
argument of the strong coupling on the r.h.s. of 
relation~(\ref{Selfconsistency}).
Further, the mass values in
equation~(\ref{Selfconsistency}) have been taken to be the pole values.
For the quark (pole) masses we have chosen $M_t=175$~GeV, $M_b=5$~GeV
and $M_c=1.7$~GeV. The reader should note that the prescription given
above to calculate the size of the strong coupling is far from being
unique. Depending on the choice of the definition of the strong
coupling, the quark mass values or the number of loops in the QCD
beta-function, larger or smaller values for $\alpha_s$ might result.
This dependence on the prescription is particularly strong for
the charm system\footnote{
As an example, using the two-loop QCD beta-function results in
$\alpha_s=0.13-0.17$, $0.27-0.44$ and $0.38-0.76$ for the top, bottom 
charm system, respectively. At this point it is clearly obvious that
the situation for the charm system is rather hopeless as far as the 
question of perturbativity is concerned.
}. 
As a consequence the theoretical
uncertainties quoted in this work should be more understood as good
guesses rather than strict theoretical limits.
However, we think that the ranges of the strong coupling given above
are good enough in order to illustrate the impact of the 
${\cal{O}}(C_F^2\alpha_s^2)$ Darwin corrections in particular for
$t\bar t$ production in the threshold region. We also want
to emphasize that our conclusions for the 
perturbativity of the different heavy quark systems do not depend on
different prescriptions for the strong coupling.
\par
\begin{table}[h]
\vskip 7mm
\begin{center}
\begin{tabular}{|c||c|c|c|c|c|c|} \hline
$n$ & $1$ &  $2$ & $3$ & $4$ & $\infty$ \\ \hline\hline
\parbox{1cm}{\vskip 2mm $\Delta_{\Psi,n}^{t\bar t}$ \vskip 2mm} & 
   $-0.05\,/\,-0.04$ & $-0.04\,/\,-0.03$ & $-0.03\,/\,-0.02$ & 
   $-0.03\,/\,-0.02$ & $-0.02\,/\,-0.02$  \\ \hline 
\parbox{1cm}{\vskip 2mm $\Delta_{\Psi,n}^{b\bar b}$ \vskip 2mm} & 
   $-0.20\,/\,-0.11$ & $-0.09\,/\,-0.06$ & $-0.06\,/\,-0.05$ & 
   $-0.05\,/\,-0.04$ & $-0.02\,/\,+0.01$  \\ \hline 
\parbox{1cm}{\vskip 2mm $\Delta_{\Psi,n}^{c\bar c}$ \vskip 2mm} & 
   $-0.34\,/\,-0.17$ & $-0.10\,/\,-0.09$ & $-0.06\,/\,-0.01$ & 
   $-0.05\,/\,+0.03$ & $-0.01\,/\,+0.15$  \\ \hline 
\end{tabular}
\caption{\label{Tablewavefunctions} 
  The relative ${\cal{O}}(C_F^2\alpha_s^2)$ Darwin 
  corrections to the moduli squared of the $l=0$ bound state wave
  functions $\Delta_{\Psi,n}$ are given for the
  $t\bar t$, $b\bar b$ and $c\bar c$ system, respectively.
  Displayed are the smallest and largest values for the range of
  $\alpha_s$ values given below eq.~(\ref{Selfconsistency}) for
  the radial quantum numbers $n=1,2,3,4$ and for $n\to\infty$. 
}
\end{center}
\vskip 3mm
\end{table}
In Table~\ref{Tablewavefunctions} the smallest and largest values
for the relative ${\cal{O}}(C_F^2\alpha_s^2)$ Darwin 
corrections to the moduli squared of the $l=0$ bound state wave
functions $\Delta_{\Psi,n}\equiv
\delta\,|\Psi_n^{\mbox{\tiny QCD}}(0)|^2_{Dar}/
|\Psi_n^{\mbox{\tiny QCD}}(0)|^2$ for the different heavy
quark systems are displayed for the ground
states ($n=1$) and the first three radial excited states ($n=2,3,4$),
employing the ranges for the strong coupling as given below 
eq.~(\ref{Selfconsistency}).
For illustration the corresponding value for ($n\to\infty$) is also
presented. The absolute values of the corrections to the ground states
amount to $4\%-5\%$ for the $t\bar t$, $11\%-20\%$ for $b\bar b$ and 
$17\%-34\%$ for the $c\bar c$ system. It is an interesting fact that 
for the $b\bar b$ and $c\bar c$ systems the size of the
corrections is rapidly  decreasing for higher excited states. In particular,
the sensitivity of  the corrections to the different values of $\alpha_s$
seems to be surprisingly small for the excited states in the $b\bar b$
and $c\bar c$ systems. We will come back to this point later.
\par
\begin{figure}[t]
 \begin{center}
\epsfxsize=4cm
\leavevmode
\epsffile[220 420 420 550]{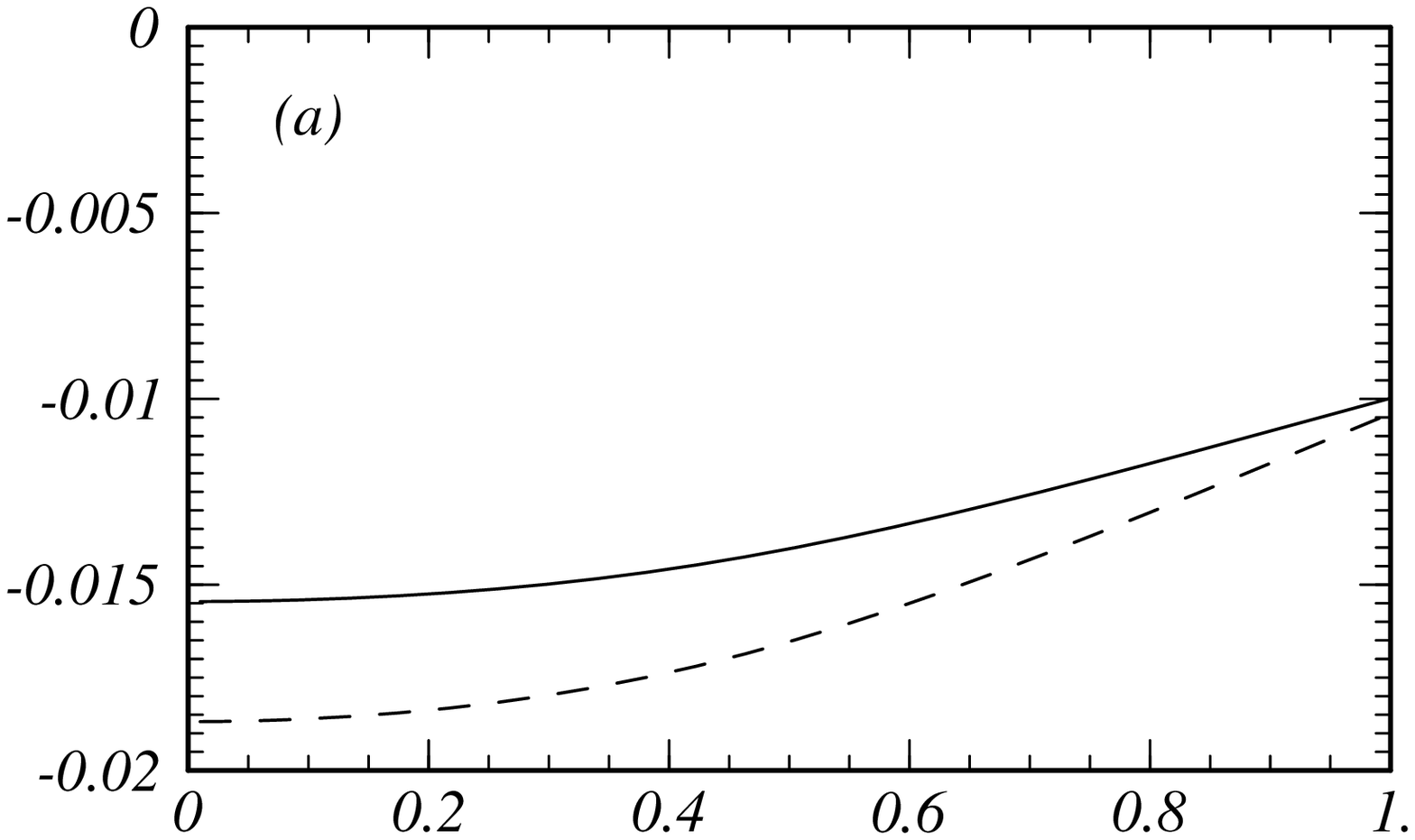}\mbox{\hspace{4mm}}\\
\vskip 3.3cm
\epsfxsize=4cm
\leavevmode
\epsffile[220 420 420 550]{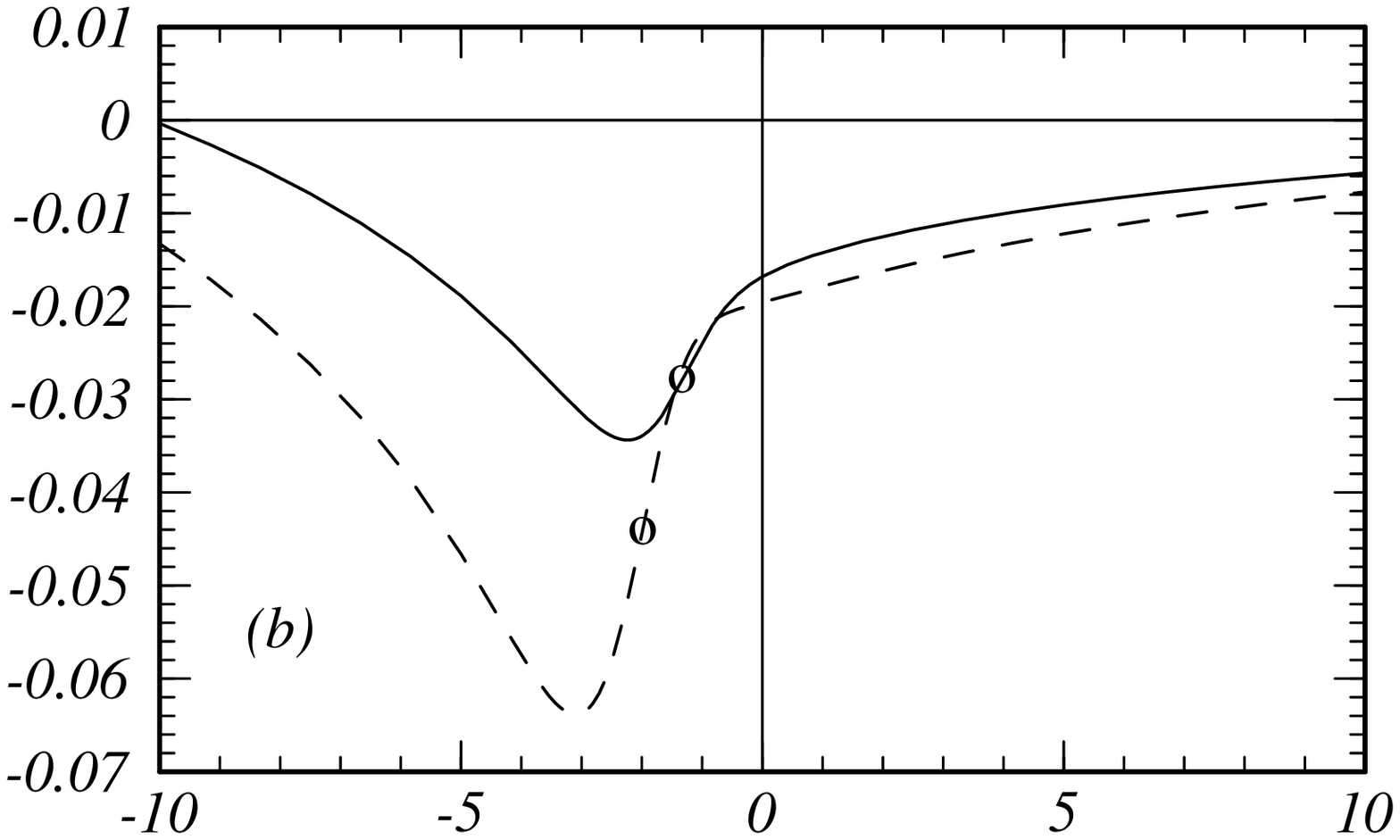}
\vskip -82mm  $\displaystyle{\bf \Delta R_{t\bar t}\mbox{\hspace{12.5cm}}}$
\vskip  5mm  $\displaystyle{\mbox{\hspace{3cm}}\bf \Gamma_t=0}$
\vskip  29mm  $\displaystyle{\mbox{\hspace{10.5cm}}\bf\frac{\beta}{C_F\,\alpha_s}}$
\vskip  7mm  $\displaystyle{\bf \Delta R_{t\bar t}\mbox{\hspace{12.5cm}}}$
\vskip  20mm  $\displaystyle{\mbox{\hspace{4.3cm}}
                                \bf \Gamma_t=1.5\,\mbox{GeV}}$
\vskip  16mm  $\displaystyle{\mbox{\hspace{10.5cm}}\bf E (\mbox{GeV})}$
\vskip 2mm
 \caption{\label{Figcrosssection1} 
 The relative ${\cal{O}}(C_F^2\alpha_s^2)$ Darwin 
 corrections to the $t\bar t$ production cross section in the 
 threshold region for the cases $\alpha_s=0.13$ (solid lines) 
 and $0.16$ (dashed lines) for stable (a) and unstable (b) top quarks.
 The circles in Figure (b) indicate the location of the $1S$ Coulomb
 energy level.}
 \end{center}
\end{figure}
\begin{figure}[t]
 \begin{center}
\epsfxsize=4cm
\leavevmode
\epsffile[220 420 420 550]{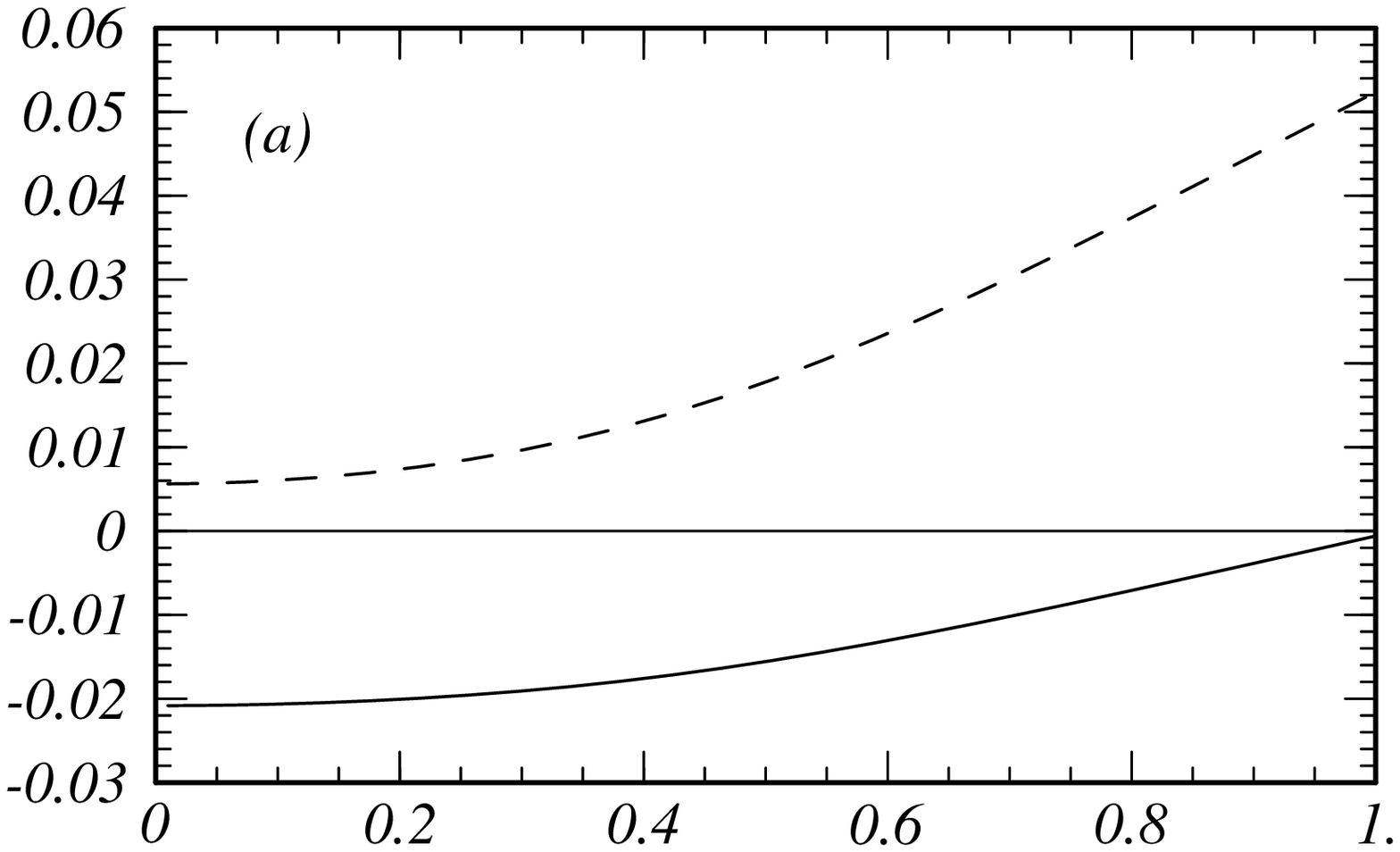}\\
\vskip 3.3cm
\epsfxsize=4cm
\leavevmode
\epsffile[220 420 420 550]{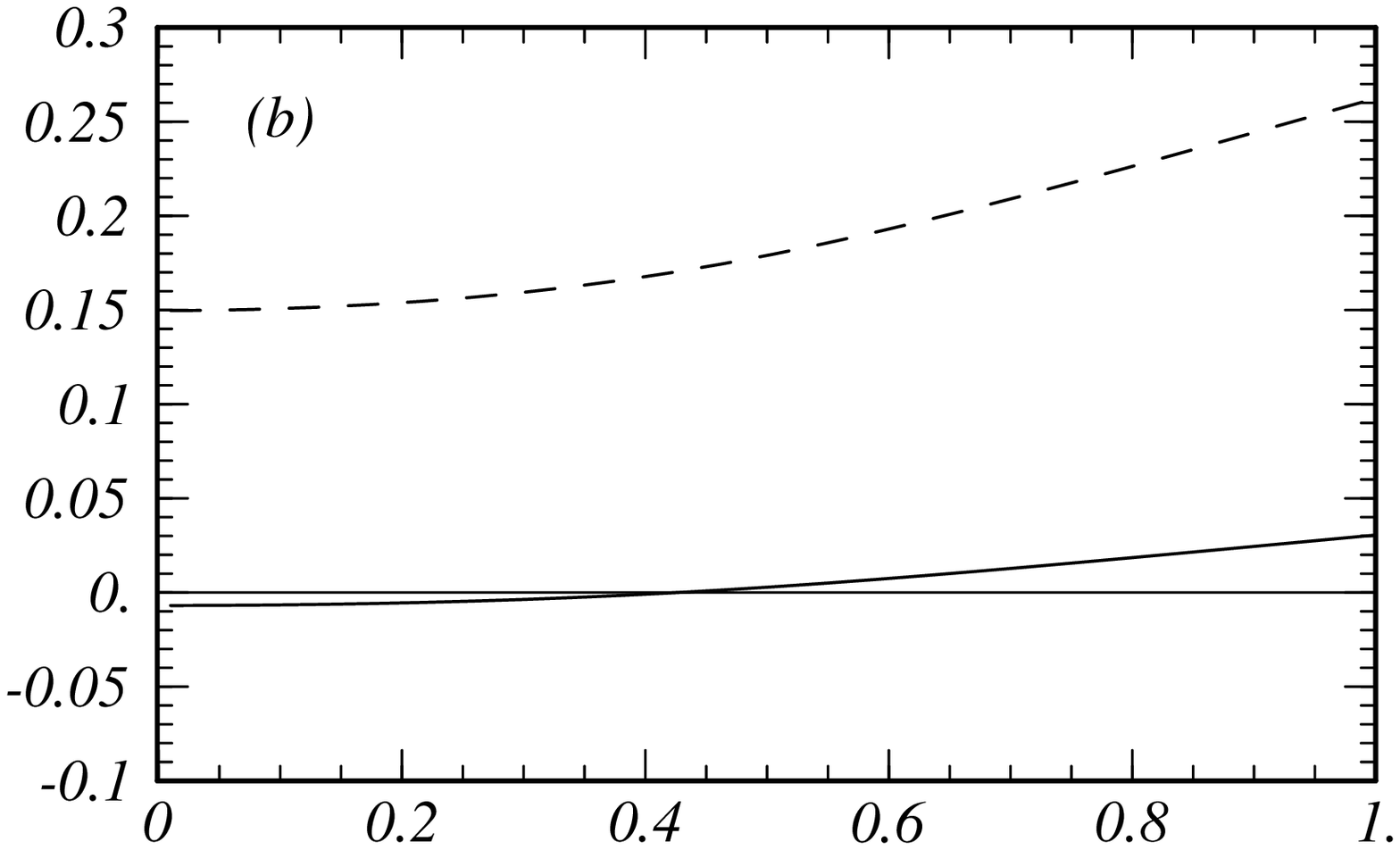}
\vskip -82mm  $\displaystyle{\bf \Delta R_{b\bar b}\mbox{\hspace{12.5cm}}}$
\vskip  39.5mm  $\displaystyle{\mbox{\hspace{10.5cm}}\bf\frac{\beta}{C_F\,\alpha_s}}$
\vskip  7mm  $\displaystyle{\bf \Delta R_{c\bar c}\mbox{\hspace{12.7cm}}}$
\vskip  39.7mm  $\displaystyle{\mbox{\hspace{10.5cm}}\bf\frac{\beta}{C_F\,\alpha_s}}$
\vskip 1mm
 \caption{\label{Figcrosssection2} 
 The relative ${\cal{O}}(C_F^2\alpha_s^2)$ Darwin 
 corrections to the $b\bar b$ (a) and $c\bar c$ (b) 
 production cross section
 in the kinematical region $0<\beta< C_F\alpha_s$ above threshold.
 The solid line corresponds to $\alpha_s=0.25$ ($0.34$) and the 
 dashed line to $\alpha_s=0.38$ ($0.59$) for the case of
 $b\bar b$ ($c\bar c$) production.}
 \end{center}
\end{figure}
In Fig.~\ref{Figcrosssection1}a and \ref{Figcrosssection2}a,b 
the relative ${\cal{O}}(C_F^2\alpha_s^2)$
Darwin corrections to the (stable) quark-antiquark production cross section 
$\Delta R_{Q\bar Q}\equiv \delta R^{Q\bar Q}_{Thr, Dar}/R^{Q\bar Q}_{Thr}$
are displayed above the threshold point 
for the three heavy quark systems in the range $0 < \beta <
C_F\,\alpha_s$. (For $t\bar t$ production this corresponds to the
energy range $0 < E < 5$~$(8)$~GeV for $\alpha_s=0.13$~$(0.16)$) 
The solid (dashed) lines correspond
to the lower (upper) $\alpha_s$ value given below 
eq.~(\ref{Selfconsistency}). For the $t\bar t$ system the size of the
relative corrections is quite stable between $-1.9\%$ and $-1.0\%$
with the tendency to decrease in magnitude for larger 
velocities. It is striking
that the dependence of the corrections on the changes in the
$\alpha_s$ value is weaker for larger velocities ($0.3\%$ for
$\beta=0$ and $0.05\%$ for $\beta=C_F\alpha_s$). For the $b\bar b$
system the corrections vary between $-2\%$ (lower value) and 
$+5\%$ (upper value)  where the larger
values occurs for larger velocities. In contrast to the top system 
the dependence of the corrections on the changes in the
$\alpha_s$ value ($3\%$ for $\beta=0$ and $5\%$ for
$\beta=C_F\alpha_s$) 
increases for larger velocities. This indicates that the
perturbative approach employed in this work works better for the $t\bar t$
than for the $b\bar b$ system. For the $c\bar c$ system, on the other
hand, the dependence on the changes in $\alpha_s$ are
tremendous. Depending on the size of the coupling the corrections vary
from $-1\%$ to $+15\%$ for $\beta=0$ up to $+3\%$ to $+26\%$ for
$\beta=C_F\alpha_s$, drawing a rather uncomfortable picture for the
perturbativity in the charm system. 
For the case of $t\bar t$ production we have also plotted the corrections 
for a finite width $\Gamma_t=1.5$~GeV (see Fig.~\ref{Figcrosssection1}b)
in the energy range $-10$~GeV~$< E < +10$~GeV 
in order to demonstrate the impact of the 
${\cal{O}}(C_F^2\alpha_s^2)$ Darwin corrections in the presence of
the large top width. This has been achieved by the naive replacement
$E\to E+i\,\Gamma_t$ in eqs.~(\ref{Darwincrosssection}) and 
(\ref{Sommerfeldfactor}).
We want to mention that the inclusion of a finite width by this naive
procedure does not represent a consistent treatment at the 
${\cal{O}}(\alpha_s^2)$ accuracy level. However, we find that this
approach is justified here in order to demonstrate that the typical
size of the ${\cal{O}}(C_F^2\alpha_s^2)$ Darwin corrections is not
altered if the top quark width is taken into account. In this case the
relative ${\cal{O}}(C_F^2\alpha_s^2)$ Darwin corrections amount to
$-6\%$ to $-2\%$ around the $1S$ peak and to $-2\%$ to $-1\%$ for higher
energies. For a rigorous
treatment of the corrections due to the off-shellness of the top quark
we refer the reader to~\cite{Modritsch1} 
and references therein.
\par
Although the  ${\cal{O}}(C_F^2\alpha_s^2)$ Darwin corrections
discussed above represent only a small part of the full
${\cal{O}}(\alpha_s^2)$ corrections, we believe that their size can be
taken as an order of magnitude estimate for the sum of all
${\cal{O}}(\alpha_s^2)$ corrections. We therefore have to face the
questions whether or how far a perturbative expansion in the strong
coupling in the threshold regime
makes sense. Because we take the position that one should not
automatically reject the possibility of a perturbative treatment
of long-distance effects,
we think that the ${\cal{O}}(C_F^2\alpha_s^2)$ Darwin corrections
determined in this work provide us with important hints toward an
acceptable answer to this fundamental question from the point of view
of perturbation theory itself. There is no doubt that perturbation 
theory in the strong coupling is still viable for the $t\bar t$
system. It has been shown in~\cite{Fadin1} by using more general arguments
that the large top mass and width serve as a screening device which
protects the $t\bar t$ properties in the threshold region from the
influence of non-perturbative effects making the $t\bar t$ system into
the ``hydrogen atom of the strong interaction''. Thus a perturbative
treatment of the $t\bar t$ system should exhibit an excellent
convergence. This is consistent with the
observations from the previous discussions showing that the  
${\cal{O}}(C_F^2\alpha_s^2)$ Darwin corrections for the top system
are at the level of a few percent for the most of the threshold
region\footnote{
A comparison of the size of the ${\cal{O}}(C_F^2\alpha_s^2)$ Darwin
corrections with the  ${\cal{O}}(C_F\alpha_s)$ corrections from
the $(1-4 C_F\alpha_s/\pi)$ suppression factor is slightly
misleading in this context because the latter represents a pure 
short-distance contribution. Therefore the
${\cal{O}}(C_F\alpha_s)$ correction should not be included in a
discussion on the convergence in the perturbative 
description of long-distance
corrections. However, for the convenience of the reader, the size of
the large ${\cal{O}}(C_F\alpha_s)$ corrections shall also be given. 
It has been shown in~\cite{Voloshin1,Hoang1,Hoang2} 
in a two-loop analysis
that the scale in the strong coupling of the
${\cal{O}}(C_F\alpha_s)$ suppression factor is $e^{-11/24}\,M_Q$
in the $\overline{\mbox{MS}}$ scheme. This results in
$-4\,C_F\,\alpha_s/\pi=-20\%$, $-41\%$ and $-64\%$ for the
top, bottom and charm systems, respectively, using the one-loop 
QCD beta-function, the pole mass values given below 
eq.~(\ref{Selfconsistency}) and 
$\alpha_s(M_z=91.187\,\mbox{GeV})=0.125$.
}. 
This, on the other hand, allows us to conclude that the
theoretical uncertainty of all present analyses for the total 
$t\bar t$ cross section in the threshold region is at the few percent 
level, because no full ${\cal{O}}(\alpha_s^2)$ treatment has ever been
accomplished there. Further, the theoretical uncertainty of such a
complete analysis  would then be roughly one to two percent around the 
$1S$ peak and below several per mille for higher energies.
This can be estimated by taking the $\alpha_s$ values presented
below~eq.~(\ref{Selfconsistency}) cubed (assuming that no scales 
lower than $C_F\alpha_s M_Q$ are relevant for the corrections
beyond the ${\cal{O}}(\alpha_s^2)$ accuracy level) and by observing
the sensitivity of the ${\cal{O}}(C_F^2\alpha_s^2)$ Darwin
corrections to changes in the values of the strong coupling
(see Fig.~\ref{Figcrosssection1}b).
To achieve an accuracy much below the percent level at the $1S$ peak 
a more rigorous treatment of the scale in the strong coupling
governing the  ${\cal{O}}(\alpha_s^2)$ corrections would be needed, 
{\it i.e.} an ${\cal{O}}(\alpha_s^3)$ calculation.
\par
As far as the $b\bar b$ system is concerned, the situation is worse
than for the $t\bar t$ system. It has been shown in a number of 
classical papers~\cite{Shifman1,Leutwyler1,Voloshin3} that a proper 
theoretical description of the bottom system can only be achieved by 
taking into account non-perturbative corrections, which cannot be 
calculated from first principles in QCD. On the other hand, it has 
been demonstrated in~\cite{Yndurain1} 
that a quite acceptable ``parameter-free'' description 
of the (S-wave, $l=0$) $b\bar b$ bound states with low radial
excitation is possible
by using perturbative calculations supplemented by
non-perturbative contributions in the form of the quark or the 
gluon condensates. However, the latter analyses (as far as corrections
to the moduli squared of the wave functions at the origin and to 
the cross section above
threshold are concerned) were essentially based on formulae 
including only the effects of the one-loop running of the strong
coupling and the global ${\cal{O}}(\alpha_s)$
correction factor  $(1-4 C_F\alpha_s/\pi)$. The question whether 
the ${\cal{O}}(\alpha_s^2)$  perturbative corrections lead to a still
converging series 
was not addressed explicitly. Equipped with the results for the
${\cal{O}}(C_F^2\alpha_s^2)$ Darwin corrections, we are able to draw
a rough picture concerning the latter question for the case of
the moduli squared of the $l=0$ bound state wave functions at the origin. 
For the ground
state the ${\cal{O}}(\alpha_s^2)$ corrections should be between $10\%$ and
$20\%$ (where the actual sign of the corrections can only be
determined by a complete ${\cal{O}}(\alpha_s^2)$ analysis)  with 
theoretical uncertainties of order $\pm 5\%$ coming from
the ignorance of the actual scale of the strong coupling and other
corrections beyond the   ${\cal{O}}(\alpha_s^2)$ level. This does not
represent an overwhelming convergence, but it is acceptable compared to
the precision of experimental measurements~\cite{Experiment1} 
and it indicates
that an actual determination of all ${\cal{O}}(\alpha_s^2)$ correction
would lead to a considerable improvement of the precision of the
theoretical description. It 
is remarkable that the ${\cal{O}}(C_F^2\alpha_s^2)$ Darwin corrections
seem to indicate that the size of the ${\cal{O}}(\alpha_s^2)$ corrections
including their sensitivity to
changes in the value of the strong coupling is much smaller for higher
excited states (see Table ~\ref{Tablewavefunctions}). Here, however,
non-perturbative contributions get more and more out of
control~\cite{Leutwyler1,Voloshin3} and a 
complete  ${\cal{O}}(\alpha_s^2)$ analysis is
therefore necessary to give a trustworthy interpretation of this
phenomenon. The latter remark is also true for the $c\bar c$ system.
\par
Finally, we also want to mention the $c\bar c$ system. In view of the 
${\cal{O}}(C_F^2\alpha_s^2)$ Darwin corrections, we can expect the
size of the complete ${\cal{O}}(\alpha_s^2)$ corrections to the
modulus squared of the ground state wave function at the origin
to be at least at the level of $15\%$ to $35\%$ 
with theoretical errors which might be almost as large
as the size of the ${\cal{O}}(\alpha_s^2)$ corrections themselves.
(Again we can estimate the size of the corrections beyond the
${\cal{O}}(\alpha_s^2)$ level by taking the long-distance $\alpha_s$
values given below eq.~(\ref{Selfconsistency}) cubed.)
It is evident that in the case
of the $c\bar c$ system the limits of perturbation theory are reached 
or even exceeded. Even with a complete determination of all
${\cal{O}}(\alpha_s^2)$ corrections the theoretical uncertainties 
would not decrease considerably, which is obviously a consequence of
the large size of the strong coupling. We therefore conclude that it
will be extremely difficult (if not impossible) to achieve a 
perturbation theory based theoretical description for
the charm system with uncertainties lower than several times $10\%$ if 
there is no (unforeseen) cancellation among different types of corrections.
\par
To conclude this section there is a remark in order: for the
calculations of the ${\cal{O}}(C_F^2\alpha_s^2)$ Darwin corrections
we used the renormalized Green function at zero distances,
eq.~(\ref{Greenfunctionrenormalized}), without any further
explanation. This
is slightly misleading because it implies that the  
${\cal{O}}(C_F^2\alpha_s^2)$ Darwin corrections to wave
functions and cross sections can be uniquely separated from all the
other ${\cal{O}}(C_F^2\alpha_s^2)$ corrections. As far as the
$\ln\alpha_s$ contribution and the digamma
term are concerned this is definitely true, but this is not the
case for the constant terms. This is a consequence of the divergences
which arise during the calculations and which have to be renormalized. 
The use of our renormalized zero-distance Green function 
represents one possible way to
achieve this renormalization. Nevertheless we think that our approach
is justified in order to illustrate the possible size of the 
complete ${\cal{O}}(\alpha_s^2)$ corrections. This view is also
supported by the explicit results for all 
${\cal{O}}(C_F^2\alpha_s^2)$ corrections to the $l=0$ wave functions 
at the origin and the cross section, which will be published
shortly. However, we want to emphasize that the latter considerations
do not affect the validity of the expressions for the vacuum
polarization function presented in Sections~\ref{SectionCalculation}
and \ref{SectionAnalysis}. There, all constants are correct due to
proper matching to the well established one- and two-loop expressions
$\Pi^{(1)}$ and $\Pi^{(2)}$, eqs.~(\ref{Pi1loopexpanded}) and 
(\ref{Pi2loopexpanded}).
\par
\vspace{.5cm}
\section{Comment on Threshold Effects far from the \\Threshold Region}
\label{SectionThreshold}
In this section we want to comment on the use and the interpretation
of the expression of the QED vacuum polarization function valid for all
energies to ${\cal{O}}(\alpha^2)$ accuracy, eq.~(\ref{Piallenergies}).
\par
We have shown in Section~\ref{SectionAnalysis} that the function $A$,
which represents the resummed expression for diagrams with the 
instantaneous Coulomb exchange of two and more longitudinal polarized 
photons (in Coulomb gauge), see eq.~(\ref{Adefinition}), essentially
has to be added to the one- and two-loop expressions for the vacuum
polarization function in order to achieve ${\cal{O}}(\alpha^2)$
accuracy in the threshold region $|\beta|\lsim\alpha$. Far from
the threshold regime, however, $A$ represents contributions of order 
$\alpha^3$ and higher and therefore is irrelevant. 
This is what we mean  by using
the term ``valid for all energies to ${\cal{O}}(\alpha^2)$ accuracy''.
-- But not more!
\par
At this point the reader might be tempted to apply
formula~(\ref{Piallenergies}), as it stands, for an energy regime far
from the threshold in the believe $A$ would represent higher-order
information which should improve the accuracy of the one- and two-loop
expressions calculated in the framework of conventional perturbation
theory. Let us illustrate such a scenario for the energy regime where
$q^2$ is close to zero. 
In this kinematic region, formula~(\ref{Piallenergies})
can be expanded in terms of small $q^2$. Taking into account only the
first non-vanishing contributions in $q^2/M^2$  and including only
contribution up to ${\cal{O}}(\alpha^3)$ the result reads
\begin{equation}
\Pi^{{\cal{O}}(\alpha^2)}_{\mbox{\tiny QED}}(q^2) \,
\stackrel{q^2\to 0}{=}  \, 
\Big(\frac{\alpha}{\pi}\Big)\,\frac{1}{15}\,\frac{q^2}{M^2} \, + \,
\Big(\frac{\alpha}{\pi}\Big)^2\,\frac{41}{162}\,\frac{q^2}{M^2} \, + \,
\alpha^3\,\frac{\pi^2}{48}\sqrt{\frac{q^2}{M^2}} \, + \, 
{\cal{O}}(\alpha^4)
\,,
\label{Pinaiveq20}
\end{equation}
where the numerical coefficient of the ${\cal{O}}(\alpha^3)$ coefficient
is $\pi^2/48=0.21$. The corresponding multi-loop expression including
also the first non-vanishing three-loop coefficient~\cite{Chet2} reads
\begin{eqnarray}
\Pi^{\mbox{\tiny 3 loop}}_{\mbox{\tiny QED}}(q^2)
& \stackrel{q^2\to 0}{=}  &
\Big(\frac{\alpha}{\pi}\Big)\,\frac{1}{15}\,\frac{q^2}{M^2} \, + \,
\Big(\frac{\alpha}{\pi}\Big)^2\,\frac{41}{162}\,\frac{q^2}{M^2}
\nonumber\\[2mm] & & +
\Big(\frac{\alpha}{\pi}\Big)^3\,
\bigg[\, -\frac{8687}{13824} + \frac{{{\pi }^2}}{3}\,
     \left( \frac{1}{8} - \frac{1}{5}\,\ln 2 \right)  + 
    \frac{22781}{27648}\,\zeta_3 
\,\bigg] \,\frac{q^2}{M^2} \, + \, 
{\cal{O}}(\alpha^4)
\,.
\label{Picorrectq20}
\end{eqnarray}  
The numerical value of the constant term in the brackets is
$0.32$. It is evident that the ${\cal{O}}(\alpha^3)$ contributions
which come from $\Pi^{{\cal{O}}(\alpha^2)}_{\mbox{\tiny QED}}$ and
therefore contain information on the formation of positronium bound states
are much larger than the three-loop contributions. The ratio between
the former  ${\cal{O}}(\alpha^3)$ contributions and the three-loop
result even diverges for $q^2\to 0$. The overall conclusion of this
scenario would be that threshold (and therefore long-distance) effects
dominate not only in the threshold regime but also the energy
region $|q^2|\ll 4 M^2$. This is obviously wrong! 
The ``threshold effects'' in
eq.~(\ref{Pinaiveq20}) contradict the 
Appelquist-Carrazone theorem~\cite{Appelquist1} and
even represent contributions 
non-analytic at $q^2=0$. The solution to this
apparent paradox is that $\Pi^{{\cal{O}}(\alpha^2)}_{\mbox{\tiny
QED}}$ only describes the vacuum polarization function to
${\cal{O}}(\alpha^2)$ accuracy. All contributions of order $\alpha^3$
or higher have to ignored and do not represent proper higher-order
contributions. This means that the contributions of the function $A$
are necessary to achieve ${\cal{O}}(\alpha^2)$ accuracy in the
threshold region, but should be neglected if the vacuum polarization
function has to evaluated far from the threshold point.
\par
To make the latter point more explicit, let us imagine that the
analytical form of the complete three-loop contributions to the vacuum
polarization function would be known for all energies (in the same
sense as the they are known for the one- and two-loop contributions,
$\Pi^{(1)}$ and $\Pi^{(2)}$). We then could try to determine the
expression for the vacuum polarization function valid to
${\cal{O}}(\alpha^3)$ accuracy for all energies in the same way as we
have determined $\Pi^{{\cal{O}}(\alpha^2)}_{\mbox{\tiny QED}}$, which
is valid  to ${\cal{O}}(\alpha^2)$ accuracy for all energies. This would
be achieved by matching the three-loop expression for the vacuum
polarization function to the corresponding ${\cal{O}}(\alpha^3)$
formula calculated in NRQED in the same way as presented in 
Section~\ref{SectionCalculation}. The vacuum polarization function
valid to ${\cal{O}}(\alpha^3)$ accuracy for all energies would then
have the form\footnote{
In eq.~(\ref{Pi3allenergies}) $\Delta(\alpha,\beta)$ denotes
the ${\cal{O}}(\alpha^3)$ NRQED contributions, including the necessary
subtractions in order to avoid double-counting. The actual form of
these contributions is irrelevant here because we only want to discuss
the large ${\cal{O}}(\alpha^3)$ contributions in
eq.~(\ref{Pinaiveq20}). However, it is straightforward to see that
$\Delta$ contains terms of order $\alpha^3$ in the threshold
regime, but is of order $\alpha^4$ far from the threshold
point. 
}
\begin{eqnarray}
\Pi_{\mbox{\tiny QED}}^{{\cal{O}}(\alpha^3)}(q^2) &  = &
\Big(\frac{\alpha}{\pi}\Big)\,\Pi^{(1)}(q^2) \, + \,
\Big(\frac{\alpha}{\pi}\Big)^2\,\Pi^{(2)}(q^2) \, + \,
\Big(\frac{\alpha}{\pi}\Big)^3\,\Pi^{(3)}(q^2) 
\nonumber\\[2mm] & &
+ \, A(\alpha,\beta) - 
   \alpha^3\,\bigg[\,i\,\frac{\pi^2}{24\,\beta}\,\bigg]
\, + \, \Delta(\alpha,\beta)
\,.
\label{Pi3allenergies}
\end{eqnarray}
In the second line of eq.~(\ref{Pi3allenergies}) the contribution
$\alpha^3\,[\,i\,\frac{\pi^2}{24\,\beta}\,]$ has to be subtracted in
order to avoid double counting in the threshold regime  since
\begin{equation}
\Big(\frac{\alpha}{\pi}\Big)^3\,
\Pi^{(3)}(q^2) \, \stackrel{|\beta|\ll 1}{=} \,
\alpha^3\,\bigg[\,
\,i\,\frac{\pi^2}{24\,\beta}
\,\bigg]
\, + \, {\cal{O}}(\beta^0)
\,.
\label{Pi3loopexpanded}
\end{equation}
It is therefore clear that far from threshold the second line of
eq.~(\ref{Pi3allenergies}) only contains contributions of order 
$\alpha^4$ and higher (see eq.~(\ref{Asmalla})). 
Expanding now $\Pi_{\mbox{\tiny
QED}}^{{\cal{O}}(\alpha^3)}$ for small values of $q^2$ would give a 
result identical to the three-loop expression, eq.~(\ref{Picorrectq20}).
The large non-analytical ${\cal{O}}(\alpha^3)$ contribution which 
appeared in eq.~(\ref{Pinaiveq20}) would be gone. It is obvious that
this large contribution originates from the leading non-vanishing term of
$\Pi^{(3)}$ in an expansion for $|\beta|\ll 1$ evaluated for small $q^2$.
These contributions survive in $\Pi_{\mbox{\tiny
QED}}^{{\cal{O}}(\alpha^2)}$, eq.~(\ref{Piallenergies}), but are
cancelled in $\Pi_{\mbox{\tiny
QED}}^{{\cal{O}}(\alpha^3)}$, eq.~(\ref{Pi3allenergies}).
Using the same line of arguments it can easily be shown that all
contributions of the function $A$ would be cancelled if formulae for
the vacuum polarization function with successively 
higher accuracy would be determined. 
\par
The physical picture behind this cancellation can be drawn as follows:
the contributions in function $A$ are generated by vacuum polarization
diagrams with the instantaneous Coulomb exchange of two and more 
longitudinal photons, where the latter are defined in the Coulomb
gauge. In the threshold region the exchange of these longitudinal
photons represents the dominant effect, whereas all the other
interactions, for simplicity reasons called ``transverse'' in the 
following, can be neglected in a first approximation.
Although this approach is obviously not
gauge invariant from the point of view of full quantum electrodynamics,
the violation of gauge invariance is vanishing in the non-relativistic
limit. This is not true, however, far from the threshold point.
There, contributions from longitudinal and transverse photons
are equally important. Their individual sizes are extraordinarily large
but with different signs. Therefore, adding the transverse contributions 
to the contributions of function $A$ the greater part
of the large corrections will be cancelled off, leaving results
which can be obtained from conventional (multi-loop) perturbation
theory. This remains true at any level of accuracy.
From this picture it should be clear that neither effects from the formation
of $e^+e^-$ bound states nor from the Coulomb rescattering, if the
relative velocity of the  $e^+e^-$ pair is much smaller than the speed
of light, can ever lead to large corrections of the vacuum 
polarization function far from the threshold region. 
There, the contributions of the
function $A$ represent unphysical (and gauge non-invariant)
contributions which cannot even be used to estimate the size of the
real higher-order corrections\footnote{
If applied to QCD our conclusion is essentially equivalent to arguments
employed in~\cite{Novikov1,Voloshin4}.
}.
\par
We would like to remind the reader that the previous arguments are not
applicable if a high number of derivatives of the vacuum polarization
function below the threshold region is considered, $({\rm d}/{\rm d}
q^2)^n \Pi(q^2)$, $n\gg 1$. In the latter case threshold effects are
essential. This can be easily understood from the relation
\begin{equation}
{\cal{M}}_n(q^2) \, \equiv \, 
\bigg(M^2\,\frac{d}{d q^2}\bigg)^n\,\Pi(q^2)
\,\sim\,
M^{2n}\,\int\frac{d {q^\prime}^2}{{q^\prime}^2}\,
\frac{{\mbox{Im}} \Pi({q^\prime}^2)}{({q^\prime}^2-q^2)^n}
\,.
\end{equation}
For large $n$ and $|q^2|\ll 4 M^2$ the high-energy contributions 
in the dispersion
integration are strongly suppressed, which leads to the domination of
effects coming from the threshold region. This fact is the foundation
of QCD sum rule calculations. At this point we would like to take
the opportunity to comment on a recent publication where QCD sum rules
have been applied to extract $\alpha_s$ and
the bottom quark mass from experimental data on the $\Upsilon$
resonances~\cite{Pich1}. In this publication it is claimed that 
${\cal{O}}(\alpha_s^2)$ corrections to the moments
${\cal{M}}_n^{\mbox{\tiny QCD}}(0)$ have been calculated 
because two-loop corrections to the
$b\bar b$ production cross section have been included into the
analysis. It should be clear from the discussions of
Section~\ref{SectionQCD} that a two-loop calculation of the cross
section is not sufficient to describe the ${\cal{O}}(\alpha_s^2)$
corrections to the cross section in the threshold region. In the
analysis of~\cite{Pich1} this can be easily seen from the fact that
the removal of the two-loop contributions (after subtraction of the
corresponding leading and next-to-leading threshold contributions)
essentially has no effect on the results (see Table~4
in~\cite{Pich1}). The latter observation is taken as a ``final test on
the importance of higher-order corrections''. However, as shown in
Section~\ref{SectionQCD}, the ${\cal{O}}(\alpha_s^2)$ corrections to
the cross section in the threshold region are expected to be at the
$10\%$ to $20\%$ level and will therefore have a large impact on QCD
sum rule calculations in the large $n$ limit. The mistake in the
arguments of~\cite{Pich1} is that it is implicitly assumed that the
Sommerfeld factor, eq.~(\ref{Sommerfeldfactor}), accounts for the
resummation of all long-distance effects. Therefore all corrections
to expression~(\ref{Sommerfeldfactor}) should be calculable by
fixed-order loop calculations alone. This
is true for the ${\cal{O}}(\alpha_s)$ short-distance correction
factor $(1-4 C_F\alpha_s/\pi)$, but this is not the case for
higher-order corrections like the ${\cal{O}}(C_F^2\alpha_s^2)$ Darwin
corrections calculated in Section~\ref{SectionQCD}. This fact will be
demonstrated explicitly in a future publication, where all
${\cal{O}}(C_F^2\alpha_s^2)$ corrections to the cross section will be
presented. In~\cite{Pich1} is it also assumed that the effects of the
running of the strong coupling in the Sommerfeld factor can be
determined by insertion of the effective running coupling $\alpha_V$, 
which is related to the short-distance corrections of the
QCD potential~\cite{Fischler1,Billoire1}. We would like to emphasize
that this approach is not justified for large $n$ QCD sum rule
calculations because the important saturation effects 
are neglected in this procedure. As a consequence the calculations
presented in~\cite{Pich1} are not only not at the
${\cal{O}}(\alpha_s^2)$ accuracy level but also include a systematic
error at order $\alpha_s$ and therefore contain much larger
uncertainties than presented there.
The authors of~\cite{Pich1} finally criticize an older QCD sum rule
calculation by Voloshin~\cite{Voloshin1} on the same subject, claiming
that in~\cite{Voloshin1} the magnitude of 
higher-order corrections was
underestimated. 
In this point we agree with the authors of~\cite{Pich1}
because in~\cite{Voloshin1} it is assumed that 
${\cal{O}}(\alpha_s^2)$ corrections have ``no enhancement'' in the
large $n$ limit and therefore should be of order $1/n$. This
assumption is essentially equivalent to the statement that all
${\cal{O}}(\alpha_s^2)$ corrections to the ground state Coulomb 
wave function at the origin of a bound $b\bar b$ pair 
should vanish. We have shown
explicitly in this work that this assumption is not true by
calculating the ${\cal{O}}(C_F^2\alpha_s^2)$ Darwin corrections. The
latter corrections amount to $10\%$ to $20\%$ for the modulus squared
of the ground state wave function at the origin 
for a bound $b\bar b$ pair and are
far from being negligible. We therefore conclude that the results 
presented in~\cite{Voloshin1} actually contain 
theoretical uncertainties at the $10\%$ to $20\%$ level,
an order of magnitude  larger than claimed there.
\par
\vspace{.5cm}
\section{Summary}
\label{SectionSummary}
In this work we have used the concept of effective field theories to
calculate the ${\cal{O}}(\alpha^2)$ corrections to the QED vacuum
polarization function in the threshold region and to define a
renormalized version of the zero-distance 
Coulomb Green function. In
the framework where non-relativistic quantum mechanics is part of an
effective low energy field theory (NRQED), long-distance effects
(coming from typical momentum scales below the electron mass) are
determined completely by classical quantum mechanics calculations,
whereas short-distance contributions (coming from momentum scales
beyond the electron mass) are included via the matching procedure. For
the latter contributions multi-loop techniques (in conventional 
covariant perturbation theory) have to be employed. We have
demonstrated that the effective field theory approach represents a
highly efficient method to merge sophisticated multi-loop methods with
well-known textbook quantum mechanics  time-independent perturbation
theory. From the physical point of view this is achieved because the
effective field theory concept allows for a transparent and systematic
separation of long- and short-distance physics at any level of
precision.
For our calculations we have used a ``direct matching'' procedure
which can be applied if the multi-loop results to the quantity of
interest are at hand. This direct matching allows for a quite sloppy
treatment of UV divergences in the effective field theory, but is of
no value if calculations of quantities are intended for which no
multi-loop expressions are available.
\par
We have demonstrated the efficiency of our approach by calculating
the ${\cal{O}}(\alpha^6)$ ``vacuum polarization'' contributions to the
positronium ground state hyperfine splitting without referring back to
the Bethe-Salpeter equation, and by determining  the
${\cal{O}}(C_F^2\alpha_s^2)$ (next-to-next-to-leading order) Darwin
corrections to heavy quark-antiquark bound state wave functions at the
origin and to the heavy quark-antiquark production cross section in
$e^+e^-$ annihilation (into a virtual photon). If the  
${\cal{O}}(C_F^2\alpha_s^2)$  Darwin corrections are taken as an 
order-of-magnitude estimate for the complete (yet unknown)
${\cal{O}}(\alpha_s^2)$ corrections, the typical
${\cal{O}}(\alpha_s^2)$ corrections for the $t\bar t$ production cross
section can be expected at the few percent
level for most of the threshold region. Around the $1S$ peak they
might even amount to $5\%$. For the modulus squared of the ground
state wave function of a bound $b\bar b$ pair 
(applicable to $\Upsilon(1S)$), the 
${\cal{O}}(C_F^2\alpha_s^2)$  Darwin corrections are between
$10\%$ and $20\%$, whereas the corresponding corrections for the 
$c\bar c$ system are between $15\%$ and $35\%$. The uncertainties
arise from the ignorance of higher-order corrections, in particular
from the ignorance of the exact scale in the strong coupling.
We conclude that the determination of all ${\cal{O}}(\alpha_s^2)$
corrections would represent a considerable improvement of the present
precision of theoretical calculations to the $t\bar t$ and $b\bar b$
system in the threshold region. For the $c\bar c$ system, on the other
hand, this seems to be doubtful, a consequence of the large size of the
strong coupling.
\par
Finally, we have also discussed whether the formation of positronium
states can lead to large corrections of the QED vacuum polarization function 
far from the threshold region and came to the conclusion that such corrections 
do not exist.
\par
\vspace{.5cm}
\section*{Acknowledgement}
I am grateful to R.~F.~Lebed and A.~V.~Manohar for many useful discussions
and reading the manuscript and to J.~Kuti and  I.~Z.~Rothstein for helpful 
conversations. I would especially like to thank P.~Labelle for
enlightening discussions on the role of logarithmic terms in bound
state problems. 

\vspace{1.0cm}
%
\sloppy
\raggedright
\def\app#1#2#3{{\it Act. Phys. Pol. }{\bf B #1} (#2) #3}
\def\apa#1#2#3{{\it Act. Phys. Austr.}{\bf #1} (#2) #3}
\def\lhc{Proc. LHC Workshop, CERN 90-10}
\def\npb#1#2#3{{\it Nucl. Phys. }{\bf B #1} (#2) #3}
\def\nP#1#2#3{{\it Nucl. Phys. }{\bf #1} (#2) #3}
\def\plb#1#2#3{{\it Phys. Lett. }{\bf B #1} (#2) #3}
\def\prd#1#2#3{{\it Phys. Rev. }{\bf D #1} (#2) #3}
\def\pra#1#2#3{{\it Phys. Rev. }{\bf A #1} (#2) #3}
\def\pR#1#2#3{{\it Phys. Rev. }{\bf #1} (#2) #3}
\def\prl#1#2#3{{\it Phys. Rev. Lett. }{\bf #1} (#2) #3}
\def\prc#1#2#3{{\it Phys. Reports }{\bf #1} (#2) #3}
\def\cpc#1#2#3{{\it Comp. Phys. Commun. }{\bf #1} (#2) #3}
\def\nim#1#2#3{{\it Nucl. Inst. Meth. }{\bf #1} (#2) #3}
\def\pr#1#2#3{{\it Phys. Reports }{\bf #1} (#2) #3}
\def\sovnp#1#2#3{{\it Sov. J. Nucl. Phys. }{\bf #1} (#2) #3}
\def\sovpJ#1#2#3{{\it Sov. Phys. LETP Lett. }{\bf #1} (#2) #3}
\def\jl#1#2#3{{\it JETP Lett. }{\bf #1} (#2) #3}
\def\jet#1#2#3{{\it JETP Lett. }{\bf #1} (#2) #3}
\def\zpc#1#2#3{{\it Z. Phys. }{\bf C #1} (#2) #3}
\def\ptp#1#2#3{{\it Prog.~Theor.~Phys.~}{\bf #1} (#2) #3}
\def\nca#1#2#3{{\it Nuovo~Cim.~}{\bf #1A} (#2) #3}
\def\ap#1#2#3{{\it Ann. Phys. }{\bf #1} (#2) #3}
\def\hpa#1#2#3{{\it Helv. Phys. Acta }{\bf #1} (#2) #3}
\def\ijmpA#1#2#3{{\it Int. J. Mod. Phys. }{\bf A #1} (#2) #3}
\def\ZETF#1#2#3{{\it Zh. Eksp. Teor. Fiz. }{\bf #1} (#2) #3}
\def\jmp#1#2#3{{\it J. Math. Phys. }{\bf #1} (#2) #3}
\def\yf#1#2#3{{\it Yad. Fiz. }{\bf #1} (#2) #3}

\end{document}